\documentclass[prd,twocolumn,showpacs,showkeys,preprintnumbers,floatfix,
  nofootinbib]{revtex4-1}
\usepackage{amsfonts} 
\usepackage{amssymb} 
\usepackage{amsmath} 
\usepackage{subfigure}
\usepackage{graphicx} 
\usepackage{array} 
\usepackage{dcolumn} 
\usepackage{bm} 
\usepackage{latexsym} 
\usepackage{longtable} 
\usepackage{hyperref} 
\usepackage{textcomp}
\usepackage[usenames,dvipsnames]{xcolor}
\usepackage{mathrsfs}
\usepackage{comment}
\usepackage{rotating}
\usepackage{here}
\usepackage{float}
\usepackage{multirow}

\graphicspath{{./figs/}}

\allowdisplaybreaks

\providecommand{\wbar}[1]{\overline#1}
\providecommand{\abs}[1]{\lvert#1\rvert}

\providecommand{\mate}[3]{\langle#1\lvert#2\rvert#3\rangle}

\renewcommand{\Re}{\mathrm{Re}\,}
\renewcommand{\Im}{\mathrm{Im}\,}

\providecommand{\GeV}{\mathrm{GeV}}
\providecommand{\MeV}{\mathrm{MeV}}

\providecommand{\CL}{\nonumber\\}

\definecolor{HLBlue}{HTML}{6599FF}
\definecolor{HLOrange}{HTML}{FF6600}

\newcommand{\MSb}{\overline{\mathrm{MS}}}

\newcommand{\mr}[1]{\ensuremath{\mathrm{#1}}}

\newcommand{\ohf}{\tfrac{1}{2}}
\newcommand{\msbar}{\ensuremath{\overline{\text{MS}}}}

\providecommand{\CL}{\nonumber\\}

\newcommand{\BK}{\hat{B}_{K}}
\newcommand{\Vcb}{|V_{cb}|}
\newcommand{\Vub}{|V_{ub}|}
\newcommand{\Vus}{|V_{us}|}

\newcommand{\eps}{\varepsilon}

\newcommand{\epsK}{\varepsilon_{K}}

\newcommand{\black}[1]{#1} 

\begin{document}

\title{ Updated evaluation of $\varepsilon_K$ in the Standard Model
  with lattice QCD inputs}

\author{Jon A. Bailey}
\affiliation{
  Lattice Gauge Theory Research Center, FPRD, and CTP, \\
  Department of Physics and Astronomy,
  Seoul National University, Seoul 08826, South Korea
}

%

\author{Sunkyu Lee}
\affiliation{
  Lattice Gauge Theory Research Center, FPRD, and CTP, \\
  Department of Physics and Astronomy,
  Seoul National University, Seoul 08826, South Korea
}

\author{Weonjong Lee}
\email[E-mail: ]{wlee@snu.ac.kr}
\affiliation{
  Lattice Gauge Theory Research Center, FPRD, and CTP, \\
  Department of Physics and Astronomy,
  Seoul National University, Seoul 08826, South Korea
}

\author{Jaehoon Leem}
\affiliation{
  School of Physics,
  Korea Institute for Advanced Study (KIAS),
  Seoul 02455, South Korea 
}

\author{Sungwoo Park}
\affiliation{
  Los Alamos National Laboratory,
  Theoretical Division T-2,
  Los Alamos, New Mexico 87545, USA
}

\collaboration{SWME Collaboration}

\date{\today}

\begin{abstract}

We report a strong tension in $\epsK$ at the $4\sigma$ level between
the experimental value and the theoretical value calculated directly
from the standard model using lattice QCD inputs such as $\BK$,
$\Vcb$, $\Vus$, $\xi_0$, $\xi_2$, $\xi_\text{LD}$, $F_K$, and
$m_c$. The standard model with lattice QCD inputs describes only 70\%
of the experimental value of $\epsK$, and does not explain its
remaining 30\%. We also find that this tension disappears when we use
the inclusive value of $\Vcb$ (results of the heavy quark expansion
based on QCD sum rules) to determine $\epsK$.  This tension is highly
correlated with the present discrepancy between the exclusive and
inclusive values of $\Vcb$. In order to resolve, in part, the issue
with $\Vcb$, it would be highly desirable to have a comprehensive
re-analysis over the entire set of experimental data on the $\bar{B}
\to D^* \ell \bar{\nu}$ decays using an alternative parametrization of
the form factors, such as the BGL parametrization, and a comparison
with results of the CLN method.

\end{abstract}

\maketitle

\section{Introduction}
\label{sec:intr}

CP violation serves as a natural place to search for new physics
\cite{ Buchalla:1995vs, Buras1998:hep-ph/9806471}.
The CP violation in the neutral kaon system is, in particular,
attractive to us, because the experimental results are already
extremely precise \cite{Patrignani:2016xqp}, and lattice QCD allows us
to perform a high precision calculation in kaon physics \cite{
  Feng:2017voh}.
In this paper, we focus on the indirect CP violation parameter
$\epsK$, which we want to determine using lattice QCD inputs.

Indirect CP violation in the neutral kaon system is parametrized by
$\epsK$,
\begin{equation}
  \label{eq:epsK_def}
  \epsK 
  \equiv \frac{\mathcal{A}(K_L \to \pi\pi(I=0))} 
              {\mathcal{A}(K_S \to \pi\pi(I=0))} \,,
\end{equation}
where $K_L$ and $K_S$ are the neutral kaon states in nature, and
$I=0$ represents the isospin of the final two-pion state.
In experiment \cite{Patrignani:2016xqp},
\begin{align}
  \label{eq:epsK_exp}
  \epsK &= (2.228 \pm 0.011) \times 10^{-3} 
  \times e^{i\phi_\eps} \,,\CL
  \phi_\eps &= 43.52 \pm 0.05 {}^\circ \,.
\end{align}
Here, the $\epsK$ value represents an $\approx 0.2\%$ impurity of the
CP even eigenstate in the $K_L$ state, which contains $\approx 99.8\%$
of the CP odd eigenstate.
%

The standard model (SM) describes the CP violation using a single
phase in the CKM matrix elements.
Hence, if there exists another phase coming from new physics,
$\epsK$ is a natural place to find it, since $\epsK$ is highly
sensitive to it.
Therefore, it has been one of the top priorities in lattice QCD to
calculate $\epsK$ to the highest possible precision \cite{
  Feng:2017voh}.
%

In order to evaluate $\epsK$ directly from the SM, we need to know
18 input parameters \cite{Bailey:2015tba}.
Out of them, we can, in principle, obtain 7 parameters from lattice
QCD: $\BK$, $\Vcb$, $\xi_0$, $\xi_2$, $\xi_\text{LD}$, $\Vus$,
$m_c(m_c)$, and $F_K$.\footnote{ In this number count, $\xi_0$ and $\xi_2$
  are redundant.  We need to know only $\xi_0$, but it is possible to
  obtain $\xi_0$ from $\xi_2$ using $\eps'/\eps$. For more details,
  refer to Ref.~\cite{Bailey:2015tba}.}
During the last decade, lattice QCD has made such remarkable progress
in calculating $\BK$ that its error is only 1.3\% at present \cite{
  Aoki:2016frl}.
At present, the largest error in theoretical calculation of $\epsK$
comes from $\Vcb$ \cite{ Jang:2017ieg, Lee:2016xkb, Bailey:2016dzk,
  Bailey:2015tba}.
%

Here, we would like to report the final results to draw your attention
to the key issues.
Evaluating $\epsK$ directly from the SM with lattice QCD inputs, we
find that it has $4.2\sigma \sim 3.9\sigma$ tension with the
experimental result when we use exclusive $\Vcb$.\footnote{Here, the
  $4.2\sigma$ tension is obtained with the estimate of RBC-UKQCD
  for $\xi_\text{LD}$, while the $3.9\sigma$ tension
  is obtained with the BGI estimate. For more
  details, refer to Section \ref{ssec:xi_LD}. }
We also find that this tension disappears with inclusive $\Vcb$.
Hence, it is clear that the key issue is the input value of $\Vcb$;
the $4\sigma$ tension in $\epsK$ is highly correlated with the $3\sigma$
tension between exclusive and inclusive $\Vcb$ \cite{ Bailey:2015tba}.

At present, there are two independent methods to determine $\Vcb$:
one is the exclusive method and the other is the inclusive method.
In the exclusive method, the experimentalists use the exclusive decays
$\bar{B} \to D^{(*)} \ell \bar{\nu}$ to determine $\Vcb
\mathcal{F}(1)$, and then combine them with lattice QCD results for
the form factor $\mathcal{F}(w)$ to determine $\Vcb$ \cite{
  Amhis:2016xyh}.
In the final analysis, they also include results for $\Vcb/\Vub$
obtained by combining the LHCb results for the ratio of the branching
fractions between the $\Lambda_b \to \Lambda_c \ell \bar{\nu}$ and
$\Lambda_b \to p \ell \bar{\nu}$ decays with lattice QCD form factors
\cite{ Amhis:2016xyh}.
In the inclusive method, one use the heavy quark expansion (HQE) as
the theoretical framework to perform the data analysis on $\bar{B} \to
X_c \ell \bar{\nu}$ decay processes \cite{ Amhis:2016xyh}.
The current status of $\Vcb$ is, in units of $1.0 \times 10^{-3}$,
\begin{align}
  \text{exclusive $\Vcb$} &= 39.13 \pm 0.59 \quad
  \text{ from Ref.~\cite{Amhis:2016xyh} }
  \label{eq:ex-Vcb}
  \\
  \text{inclusive $\Vcb$} &= 41.98 \pm 0.45 \quad
  \text{ from Ref.~\cite{Amhis:2016xyh} }
  \label{eq:in-Vcb}
\end{align}
where the result in Eq.~\eqref{eq:in-Vcb} is obtained in the 1S
scheme.
The difference between \eqref{eq:ex-Vcb} and \eqref{eq:in-Vcb} is
$3.8\sigma$.
This gap between exclusive and inclusive $\Vcb$ has not been resolved
yet.
However, a number of interesting ideas have been proposed in order to
resolve this issue \cite{ Bigi:2017njr, Grinstein:2017nlq}.
We review them in Section \ref{sssec:cln-bgl} and Appendix
\ref{app:cln-bgl} when we discuss $\Vcb$.

The main goal of this paper is to present the most up-to-date results
for $\epsK$ obtained directly from the SM by using lattice QCD
and experimental inputs.
In Section \ref{sec:rev}, we review the master formula for $\epsK$ and
describe each term in detail, including the physical meaning.
In Section \ref{sec:input}, we explain how to obtain the 18 input
parameters one by one. In the case of $\Vcb$, caveats in various
methods for the form factor parametrization are addressed in some
detail.
In Section \ref{sec:result}, we present results for $\epsK$ obtained
using various combinations of input parameters.
In Section \ref{sec:conclude}, we conclude.

\section{Review of $\epsK$}
\label{sec:rev}

\subsection{Master Formula: $\epsK$}
\label{ssec:master}
In the standard model (SM), the direct CP violation parameter $\epsK$
in the neutral kaon system can be re-expressed in terms of the
well-known SM parameters as follows,
\begin{align}
  \epsK
  =& e^{i\theta} \sqrt{2}\sin{\theta} 
  \Big( C_{\eps} X_\text{SD} \hat{B}_{K} 
  + \frac{ \xi_{0} }{ \sqrt{2} } + \xi_\text{LD} \Big) \CL
   &+ \mathcal{O}(\omega \eps^\prime)
   + \mathcal{O}(\xi_0 \Gamma_2/\Gamma_1) \,.
  \label{eq:epsK_SM_0}
\end{align}
This is the master formula, and its derivation is well
explained in Ref.~\cite{Bailey:2015tba}.
Here, we use the same notation and convention as in
Ref.~\cite{Bailey:2015tba}.

\subsection{Short Distance Contribution to $\epsK$}
\label{ssec:SD}

In the master formula of Eq.~\eqref{eq:epsK_SM_0}, the dominant
leading-order effect ($\approx +107\%$) comes from the short distance
(SD) contribution proportional to $\BK$.
Here, $C_{\eps}$ is a dimensionless parameter defined as:
\begin{align}
  C_{\eps}
  &\equiv \frac{ G_{F}^{2} F_K^{2} m_{K^{0}} M_{W}^{2} }
  { 6\sqrt{2} \pi^{2} \Delta M_{K} } \cong 3.63 \times 10^{4} \,,
  \label{eq:C_eps}
\end{align}
Here, $X_\text{SD}$ represents the short distance effect from the
Inami-Lim functions \cite{ Inami1980:ProgTheorPhys.65.297}:
\begin{align}
  X_\text{SD} &\equiv
  \Im\lambda_t \Big[ \Re\lambda_c \eta_{cc} S_0(x_c)
    -\Re\lambda_t \eta_{tt} S_0(x_t) \CL & \quad - (\Re\lambda_c -
    \Re\lambda_t) \eta_{ct} S_0(x_c,x_t) \Big] 
  \label{eq:X_SD}
  \\
  &\cong 6.24 \times 10^{-8} \,,
\end{align}
where $\lambda_i = V^*_{is} V_{id}$ is a product of the CKM matrix
elements with $i = u,c,t$, and $\eta_{ij}$ with $i,j = c,t$ represent
the QCD corrections of higher order in $\alpha_s$ \cite{
  Herrlich1996:NuclPhysB.476.27}.
There exists a potential issue with poor convergence of perturbation
theory for $\eta_{cc}$ at the charm scale, which is discussed properly
in Ref.~\cite{ Bailey:2015tba}.
Here, $S_0$'s are Inami-Lim functions \cite{
  Inami1980:ProgTheorPhys.65.297} defined as
\begin{align}
  \label{eq:InamiFn}
  & S_{0}(x_i)
  = x_i \bigg[ \frac{1}{4} + \frac{9}{4(1-x_i)} - \frac{3}{2(1-x_i)^2}
    - \frac{3x_i^2 \ln x_i}{2 (1-x_i)^3}  \bigg] \,,\CL
  & S_{0}(x_i,x_j) 
  = \Bigg\{ \frac{x_i x_j}{x_i-x_j} 
     \bigg[ \frac{1}{4} + \frac{3}{2(1-x_i)}
    - \frac{3}{4(1-x_i)^2} \bigg] \ln x_i \CL
     &\hspace{5pc} + ( i \leftrightarrow j) \Bigg\}
     - \frac{3x_i x_j}{4(1-x_i)(1-x_j)} \,,
\end{align}
where $i = c,t$, $x_i = m_i^2/M_W^2$, and $m_i = m_i(m_i)$ is
the scale invariant $\MSb$ quark mass.
In $X_\text{SD}$ of Eq.~\eqref{eq:X_SD}, the $S_0(x_t)$ term from the
top-top contribution in the box diagrams describes about $+72.4\%$ of
$X_\text{SD}$, the $S_0(x_c, x_t)$ term from the top-charm
contribution takes over about $+45.4\%$ of $X_\text{SD}$, and the
$S_0(x_c)$ term from the charm-charm contribution depicts about
$-17.8\%$ of $X_\text{SD}$.

Here, the kaon bag parameter $\BK$ is defined as
\begin{align}
  \BK &\equiv B_K(\mu) b(\mu) \cong 0.76 \,,
  \\
  B_K(\mu)
  &\equiv
  \frac{ \mate{ \bar{K}^{0} }{ O_{LL}^{\Delta S = 2}(\mu) }{ K^{0} } }
       { \frac{8}{3}
         \mate{\bar{K}^0}{\bar{s} \gamma_{\mu}\gamma_5 d}{0}
         \mate{0}{\bar{s} \gamma^{\mu}\gamma_5 d}{K^0} } \CL
  &= \frac{ \mate{\bar{K}^0}{O_{LL}^{\Delta S = 2}(\mu)}{K^0} }
       { \frac{8}{3} F_K^2 m_{K^0}^2 } \,,
  \\
  O_{LL}^{\Delta S = 2}(\mu)
  &\equiv [\bar{s}\gamma_{\mu}(1-\gamma_5)d]
         [\bar{s}\gamma^{\mu}(1-\gamma_5)d] \,,  
\end{align}
where $b(\mu)$ is the renormalization group (RG) running factor to
make $\BK$ invariant with respect to the renormalization scale and
scheme:
\begin{align}
  b(\mu) &= [ \alpha_{s}^{(3)}(\mu) ]^{-2/9} K_+(\mu) \,.
\end{align}
Here, details on $K_+(\mu)$ are given in Ref.~\cite{Bailey:2015tba}.

\subsection{Long Distance Contribution to $\epsK$}
\label{ssec:LD}
There are two kinds of long distance (LD) contributions on $\epsK$:
one is the absorptive LD effect from $\xi_0$ and the other is the
dispersive LD effect from $\xi_\text{LD}$.
The absorptive LD effects are defined as
\begin{align}
  \tan{\xi_0} &\equiv \frac{\Im A_0}{\Re A_0} \,,
  \\
  \tan{\xi_2} &\equiv \frac{\Im A_2}{\Re A_2} \,.
\end{align}
They are related with each other through $\eps^\prime$:
\begin{align}
  \eps^\prime
  & \equiv 
  e^{i(\delta_2-\delta_0)} \frac{i\omega}{\sqrt{2}} 
  \Big( \tan{\xi_2} - \tan{\xi_0} \Big) 
  \nonumber \\
  & =   e^{i(\delta_2-\delta_0)} \frac{i\omega}{\sqrt{2}} 
  ( \xi_2 - \xi_0 ) + \mathcal{O}(\xi^3_i) \,.
  \label{eq:epsp}
\end{align}
The overall contribution of the $\xi_0$ term to $\epsK$ is about
$-7\%$.

The dispersive LD effect is defined as
\begin{align}
  \xi_\text{LD} &=  \frac{m^\prime_\text{LD}}{\sqrt{2} \Delta M_K} \,,
\label{eq:xiLD}
\end{align}
where
\begin{align}
  \label{eq:mLD}
  m^\prime_\text{LD}
  &= -\Im \left[ \mathcal{P}\sum_{C} 
    \frac{\mate{\wbar{K}^0}{H_\text{w}}{C} \mate{C}{H_\text{w}}{K^0}}
         {m_{K^0}-E_{C}}  \right]
  \,.
\end{align}
if the CPT invariance is well respected.
The overall contribution of the $\xi_\text{LD}$ to $\epsK$ is about
$\pm 2\%$.

\subsection{Erratum}
\label{ssec:typo}
There were two pure typos in Ref.~\cite{Bailey:2015tba}.
One typo is found in Eq.~(50) of Ref.~\cite{Bailey:2015tba}.
The correct equations for $S_0(x_i)$ and $S_0(x_i,x_j)$ are given in
Eq.~\eqref{eq:InamiFn} of this paper.
The other typo is found in Eq.~(62) of Ref.~\cite{Bailey:2015tba}.
The correct equation for $\delta m^\prime_\text{LD}$ is
\begin{align}
  \delta m^\prime_\text{LD} &= -i\frac{1}{2}\mathcal{P}\sum_{C}
  \frac{ \abs{\mate{K_0}{H_\text{w}}{C}}^2 - 
    \abs{\mate{\wbar{K}^0}{H_\text{w}}{C}}^2 }{m_{K^0}-E_{C}}
  \CL
  &= 0\,,
\end{align}
if the CPT invariance is well respected.
The $-i$ factor is missing in Ref.~\cite{Bailey:2015tba}.
In our actual calculation of $\epsK$, we used the correct equations
with no mistake, even through we introduced the above two typos in
writing up the paper of Ref.~\cite{ Bailey:2015tba}.

\section{Input Parameters}
\label{sec:input}
We need to know values of 18 parameters defined in the standard model
(SM) in order to evaluate $\epsK$ directly from the SM.
Out of the 18 parameters, we can obtain, in principle, 7 parameters
such as $\BK$, $\Vcb$, $\xi_0$, $\xi_2$, $\xi_\text{LD}$, $\Vus$,
$F_K$, and $m_c(m_c)$ directly from lattice QCD.
Here, we describe how to obtain the 18 input parameters from the
experiments and from lattice QCD results in detail.

\subsection{Wolfenstein Parameters}
\label{ssec:wp}
The CKMfitter \cite{ Charles:2004jd} and UTfit \cite{ Bona:2006ah}
collaborations provide the Wolfenstein parameters \cite{
  Winstein1993:RevModPhys.65.1113} ($\lambda$, $\bar{\rho}$,
$\bar{\eta}$) determined by the global unitarity triangle (UT) fit.
The 2017 results are summarized in Table \ref{tab:wp}.
As pointed out in Ref.~\cite{ Bailey:2015tba}, the Wolfenstein
parameters extracted by the global UT fit have unwanted correlation
with $\epsK$, because $\epsK$ is used as an input to obtain them.
Hence, in order to avoid this correlation, we take another set of the
Wolfenstein parameters determined from the angle-only-fit (AOF)
suggested in Ref.~\cite{ Bevan2013:npps241.89}.
In the AOF, $\epsK$, $\BK$, and $\Vcb$ are not used as inputs to
determine the UT apex ($\bar{\rho}$,$\bar{\eta}$).
Then, we determine $\lambda$ from $\Vus$ which is obtained from the
$K_{\ell2}$ and $K_{\ell3}$ decays using the lattice QCD results.
The Wolfenstein parameter $A$ is determined directly from $\Vcb$,
which will be discussed later in Section \ref{ssec:Vcb}.
The Wolfenstein parameters from the AOF are summarized in Table
\ref{tab:wp}.

\begin{table*}
  \caption{Wolfenstein parameters (WP). Both CKMfitter and UTfit groups use
    the global unitarity triangle fit. Here, AOF represents the angle
    only fit.}
  \label{tab:wp}
  \renewcommand{\arraystretch}{1.2}
  \begin{ruledtabular}
    \begin{tabular}{@{\quad} c @{\quad}| l c @{\qquad}| l c @{\qquad}| l c }
      WP
      & \multicolumn{2}{c|}{CKMfitter}
      & \multicolumn{2}{c|}{UTfit}
      & \multicolumn{2}{c}{AOF} \\ \hline
      $\lambda$
      & $0.22509(29)$ & \cite{Charles:2004jd}
      & $0.22497(69)$ & \cite{Bona:2006ah}
      & $0.2248(6)$   & \cite{Patrignani:2016xqp}
      \\ \hline
      $\bar{\rho}$
      & $0.1598(76)$ & \cite{Charles:2004jd}
      & $0.153(13)$  & \cite{Bona:2006ah}
      & $0.146(22)$  & \cite{Martinelli:2017}
      \\ \hline
      $\bar{\eta}$
      & $0.3499(63)$ & \cite{Charles:2004jd}
      & $0.343(11)$  & \cite{Bona:2006ah}
      & $0.333(16)$  & \cite{Martinelli:2017}
    \end{tabular}
  \end{ruledtabular}
\end{table*}

\subsection{$\BK$}
\label{ssec:bk}
In the FLAG review \cite{ Aoki:2016frl}, they present lattice QCD
results for $\BK$ with $N_f=2$, $N_f=2+1$, and $N_f= 2+1+1$.
Recent calculations of $\BK$ in lattice QCD have been done with
$N_f = 2+1+1$ dynamical quarks \cite{ Carrasco:2015pra}.
We do not prefer it for two physical reasons:
\begin{enumerate}
\item The master formula in Eq.~\eqref{eq:epsK_SM_0} is derived by
  integrating out the heavy particles including the charm quark.
  Hence, it is complicated and inconvenient to use the results for
  $\BK$ with dynamical charm quarks \cite{ Aoki:2016frl,
    Carrasco:2015pra}.
\item A proper and systematic procedure to incorporate the charm
  quark effect in the calculation of $\epsK$ is available in
  Refs.~\cite{ Christ:2012se, Bai:2014cva}.  However, this new method
  is in the stage of exploratory study and has not reached the stage
  of precision measurement yet.
\end{enumerate}
Similarly, we prefer using the results for $\BK$ with $N_f=2+1$ to
those with $N_f=2$ because they are obtained by quenching the vacuum
polarization contributions of the strange quark.

\begin{table}
  \caption{$\BK$ in lattice QCD with $N_f = 2+1$.}
  \label{tab:bk}
  \renewcommand{\arraystretch}{1.2}
  \begin{ruledtabular}
    \begin{tabular}{@{\quad} l @{\quad} c @{\quad} l @{\quad}}
      Collaboration & Ref. & $\BK$  \\ \hline
      SWME 15       & \cite{Jang:2015sla} & $0.735(5)(36)$     \\
      RBC/UKQCD 14  & \cite{Blum:2014tka} & $0.7499(24)(150)$  \\
      Laiho 11      & \cite{Laiho:2011np} & $0.7628(38)(205)$  \\
      BMW 11        & \cite{Durr:2011ap}  & $0.7727(81)(84)$  \\ \hline
      FLAG 17       & \cite{Aoki:2016frl} & $0.7625(97)$      
    \end{tabular}
  \end{ruledtabular}
\end{table}

In Table \ref{tab:bk}, we present the FLAG results for $\BK$ with
$N_f = 2+1$.
Here, they take a global average over the four data points from BMW
11 \cite{ Durr:2011ap}, Laiho 11 \cite{ Laiho:2011np}, RBC/UKQCD 14
\cite{ Blum:2014tka}, and SWME 15 \cite{ Jang:2015sla}.
The FLAG 17 in the table represents the final result for $\BK$.
Here, we use this for our evaluation of $\epsK$.

\subsection{$\Vcb$}
\label{ssec:Vcb}
In Table \ref{tab:Vcb}, we summarize updated results for both
exclusive $\Vcb$ and inclusive $\Vcb$.
Recently HFLAV reported them in Ref.~\cite{Amhis:2016xyh}.
The results for exclusive $\Vcb$ depend on the lattice QCD
calculations of form factors of Refs.~\cite{
  Bailey2014:PhysRevD.89.114504, Lattice:2015rga, Detmold:2015aaa}.
Here, when we obtain the in-combined results in Table
\ref{tab:Vcb}\;\subref{tab:in-Vcb}, we neglect the hidden correlation
of the inclusive $\Vcb$ between the kinetic scheme and the 1S scheme,
even though there must be some correlation because they share some
experimental data with each other.
Hence, we prefer using results of the 1S scheme to the results of
in-combined here.
We use the combined results (ex-combined) for the exclusive $\Vcb$ and
the results of the 1S scheme for inclusive $\Vcb$ when we evaluate
$\epsK$.

\begin{table}[h]
  \renewcommand{\arraystretch}{1.2}
  \caption{Results for $\Vcb$ in units of $1.0\times 10^{-3}$. The
    in-combined result is obtained by taking an uncorrelated weighted
    average of the two values in Table \subref{tab:in-Vcb}.}
  \label{tab:Vcb}
  \subtable[Exclusive $\Vcb$]{
    \begin{ruledtabular}
      \begin{tabular}{l l l}
        channel & value & Ref. \\ \hline
        $B\to D^* \ell \bar{\nu}$
        & $39.05(47)(58)$ & \cite{Amhis:2016xyh}\footnote{ In this analysis,
          they use the lattice QCD results for the semileptonic form factors
          in Ref.~\cite{ Bailey2014:PhysRevD.89.114504}.}  \\
        $B\to D \ell \bar{\nu}$
        & $39.18(94)(36)$ & \cite{Amhis:2016xyh}\footnote{ In this analysis,
          they use the lattice QCD results for the semileptonic form factors
          in Ref.~\cite{ Lattice:2015rga}.}  \\
        $|V_{ub}|/|V_{cb}|$
        & $0.080(4)(4)$   & \cite{Amhis:2016xyh}\footnote{ In this analysis,
          they use the lattice QCD results for the semileptonic form factors
          in Ref.~\cite{ Detmold:2015aaa}.}  \\ \hline
        ex-combined
        & $39.13(59)$     & \cite{Amhis:2016xyh} 
      \end{tabular}
    \end{ruledtabular}
    \label{tab:ex-Vcb} 
  } 
  \subtable[Inclusive $\Vcb$]{
    \begin{ruledtabular}
      \begin{tabular}{l l c}
        channel & value & Ref. \\ \hline
        kinetic scheme & $42.19(78)$ & \cite{Amhis:2016xyh} 
        \\ 
        1S scheme      & $41.98(45)$ & \cite{Amhis:2016xyh} 
        \\ \hline
        in-combined    & $42.03(39)$ & this paper 
      \end{tabular}
    \end{ruledtabular}
    \label{tab:in-Vcb} 
  } 
\end{table}

\begin{figure}[htbp]
  \hspace*{-5mm}
  \includegraphics[width=1.05\linewidth]{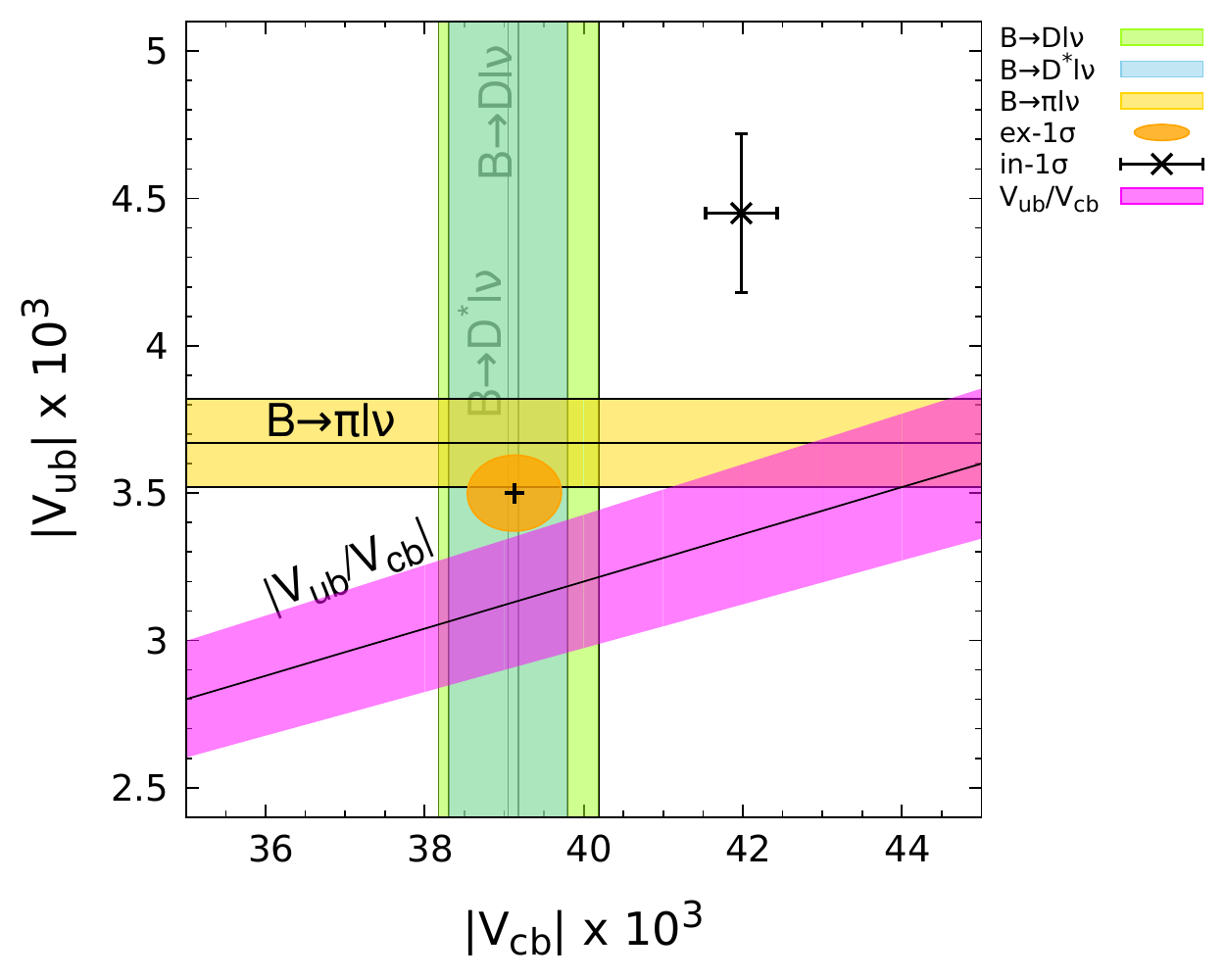}
  \caption{$\Vcb$ versus $\Vub$ in units of $1.0\times 10^{-3}$. The
    light-blue band represents $|V_{cb}|$ determined from the
    $\bar{B}\to D^* \ell \bar\nu$ decay mode. The light-green band
    represents $|V_{cb}|$ determined from the $\bar{B}\to
    D\ell\bar\nu$ decay mode. The yellow band represents $|V_{ub}|$
    determined from the $\bar{B} \to \pi\ell\bar\nu$ decay mode.  The
    magenta band represents $|V_{ub}/V_{cb}|$ determined from the LHCb
    data of the $\Lambda_b \to \Lambda_c \ell\bar\nu$ and $\Lambda_b
    \to p \ell\bar\nu$ decay modes. The orange circle represents the
    combined results for exclusive $|V_{cb}|$ and $|V_{ub}|$ from the
    $B$ meson and $\Lambda_b$ decays within $1.0\sigma$. The black
    cross (\textbf{\texttimes}) represents the inclusive $|V_{cb}|$
    and $|V_{ub}|$ from the heavy quark expansion. The details are
    given in Table \ref{tab:Vcb}.  }
  \label{fig:Vcb}
\end{figure}

In Fig.~\ref{fig:Vcb}, we present results for $\Vcb$ and $\Vub$.
The big change is that, as of Lattice 2016, the result for exclusive
$\Vcb$ from $\bar{B} \to D \ell \bar{\nu}$ was about one sigma away
from that from $\bar{B} \to D^{*} \ell \bar{\nu}$ (refer to
Ref.~\cite{ Lee:2016xkb, Bailey:2016dzk} for more details), but in
2017, they are on top of each other, as shown in Fig.~\ref{fig:Vcb}.
The 2017 results for $\bar{B} \to D^{*} \ell \bar{\nu}$ are not visibly
different, but those for $\bar{B} \to D \ell \bar{\nu}$ shift downward by
about $1\sigma$.
The difference is due to several factors acting in concert:  The 2017
results of HFLAV include all results from the $B$ factories, BABAR and BELLE,
as well as the older results from CLEO and the LEP experiments ALEPH, OPAL,
and DELPHI.
Before the results are averaged, they are rescaled by HFLAV to updated
values of the inputs, and the averages include the effects of correlations.
%

%
%
To obtain the 2017 results for exclusive $\Vcb$ (and $\Vub$), HFLAV
performed a combined fit to all results for the decays $\bar{B} \to
D^{(*)} \ell \bar{\nu}$, $B \to \pi \ell \nu$, and the
ratio of the branching fractions for $\Lambda_b \to p \ell
\bar{\nu}$ and $\Lambda_b \to \Lambda_c \ell \bar{\nu}$.
With lattice QCD results for the form factors, the $B \to \pi \ell
\nu$ decay yields $\Vub$, while the ratio of the branching fractions
of the $\Lambda_b$ decays yields $\Vub / \Vcb$.
Due to the addition of more data to the HFLAV analysis, the results
for $\Vub / \Vcb$ shift downward by about $\frac{3}{4}\sigma$ in 2017,
while those for $\Vub$ shift downward by about $0.1\sigma$.
For more details, refer to Ref.~\cite{Amhis:2016xyh}.

\subsubsection{Caveats on CLN and BGL}
\label{sssec:cln-bgl}
In order to extract a value of $\Vcb \mathcal{F}(1)$ ($\mathcal{F}(w)$
is a form factor at a recoil point $w$) from the experiment of
$\bar{B} \to D^{(*)}\ell \bar{\nu}$ we need to know the functional
form of the form factors as a function of $w$.
There have been two kinds of parametrization methods developed to do
this job: one is an HQET-dependent method, and the other is an
HQET-independent method.\footnote{Here, HQET is an abbreviation for
  the Heavy Quark Effective Theory.}
The former is a method of Caprini, Lellouch, and Neubert (CLN) in
Ref.~\cite{ Caprini:1997mu} and its sibling paper \cite{
  Caprini:1995wq}, and the latter is a method of Boyd, Grinstein, and
Lebed (BGL) in Ref.~\cite{Boyd:1997kz} and its sibling papers \cite{
  Boyd:1995sq, Boyd:1997qw, Boyd:1995tg}.\footnote{There exists a
  variant of the BGL method which is often referred to as the ``BCL
  method'' \cite{ Bourrely:2008za}. }
Both of them have been developed on top of the building blocks
designed for $K_{\ell3}$ decays in Refs.~\cite{ LingFong:1971er,
  Okubo:1971jf, Bourrely:1980gp}.

Recently, in Refs.~\cite{ Bigi:2017njr, Grinstein:2017nlq,
  Bigi:2017jbd}, they claim that the gap between the inclusive $\Vcb$
and exclusive $\Vcb$ might be explained in part by the observation of
both groups that CLN consistently underestimates the value of
exclusive $\Vcb$ compared with that of BGL.
In this claim, they refer to the numbers of HFLAV in Ref.~\cite{
  Amhis:2016xyh} which are obtained using the CLN method.
Certainly, this claim is interesting enough to deserve our full and
careful investigation on it.

Let us first describe the key points of the claim in Refs.~\cite{
  Bigi:2017njr, Grinstein:2017nlq}.
In the CLN parametrization, they introduce the form factor
$h_{A_1}(w)$, and the ratios of $R_1(w)$ and $R_2(w)$ to describe the
form factors for $\bar{B} \to D^* \ell \bar{\nu}$ decays.
Their definition (Eq.~\eqref{eq:cln-ff-def}) and detailed explanation
are given in Appendix \ref{app:cln-bgl}.
Let us directly address the problematic part in CLN.
CLN is constructed based on HQET and its perturbative application to
the slope and curvature of $h_{A_1}(w)$, $R_1(w)$, and $R_2(w)$.
CLN was originally designed to have its error in the level of about
2\% precision \cite{ Caprini:1997mu}.
At present, the trouble is that the experimental precision goes below
the 2\% level.
The lattice QCD results have precision better than that of the 2\%
level.
The typical size of errors from the slope and curvature in $R_i(w)$
obtained using the perturbation theory in HQET is about 10\% which
has a potential to cause $1\sim 2$\% errors in $\Vcb$.
They (\cite{ Bigi:2017njr, Grinstein:2017nlq}) observed that the CLN
method consistently underestimates exclusive $\Vcb$ compared with that
of the BGL method which is model-independent by construction.
Details on BGL are summarized in Appendix \ref{app:cln-bgl}.
To support their claim, they used a preliminary unfolded data of
BELLE in Ref.~\cite{ Abdesselam:2017kjf}.
In their conclusion, they recommended comprehensive reanalysis of old
experimental data used in Ref.~\cite{ Amhis:2016xyh} using the BGL
method as well as some suggestions to the lattice QCD community.

In Ref.~\cite{ Bernlochner:2017jka}, they incorporate all the
$\mathcal{O}(\Lambda_\text{QCD}/m_{c,b})$ and $\mathcal{O}(\alpha_s)$
contributions in the HQET framework into their analysis for $\Vcb$
based on the CLN method.
They find that their results for $\Vcb$ with improved precision agree
with those of HFLAV \cite{Amhis:2016xyh}.
In Ref.~\cite{ Bernlochner:2017xyx}, they use the same kind of CLN
method as in Ref.~\cite{Bernlochner:2017jka} and its variations as
well as the BGL method to determine $\Vcb$ and semi-leptonic form
factors.
In this study they find that the slope of the form factor ratio
$R_1(w)$ at zero recoil obtained using the BGL method has potentially
large deviation from heavy quark symmetry, and, in addition, has
significant tension with the preliminary lattice QCD results of
FNAL/MILC \cite{ Aviles-Casco:2017nge} and JLQCD \cite{
  Hashimoto:2018Latt, Kaneko:2018Latt}.
They point out that the tensions between the exclusive and inclusive
determinations of $\Vcb$ are far away from being considered resolved
at present.
In Ref.~\cite{ Bigi:2017jbd}, however, they claim that the conclusions
previously reached in Ref.~\cite{ Bigi:2017njr} are not changed by
taking into account heavy quark symmetry.
The extraction of $\Vcb$ using the CLN and BGL parametrizations
and preliminary BELLE data has been further investigated in
Ref.~\cite{ Jaiswal:2017rve}.

The claim in Refs.~\cite{ Bigi:2017njr, Grinstein:2017nlq} is
interesting, but far away from conclusive or decisive in that they
used only a preliminary subset of the BELLE data, and the BGL results
for the $R_1(w)$ slope has significant violation of heavy quark
symmetry and is disfavored by preliminary lattice QCD results
\cite{ Bernlochner:2017xyx}.
This issue might well be resolved one way or the other, once the next
round of comprehensive reanalysis by HFLAV on the old experimental
data used in Ref.~\cite{ Amhis:2016xyh} becomes available.
Lattice QCD calculation of semi-leptonic form factors for the $\bar{B}
\to D^{(*)} \ell \bar{\nu}$ decays at non-zero recoil will be helpful
\cite{ Hashimoto:2018Latt}.
Hence, please stay tuned for this coming update.

\subsection{$\xi_0$}
\label{ssec:xi0}
The absorptive part of long distance effects in $\epsK$ is
parametrized into $\xi_0$.
We can express $\eps'/\eps$ in terms of $\xi_0$ and $\xi_2$ as
follows,
\begin{align}
  \xi_0 &\equiv \arctan \bigg( \frac{\Im A_0}{\Re A_0} \bigg)
  = \frac{\Im A_0}{\Re A_0} + \mathcal{O}(\xi_0^3)
  \\ 
  \xi_2 &\equiv \arctan \bigg(\frac{\Im A_2}{\Re A_2} \bigg)
  = \frac{\Im A_2}{\Re A_2} + \mathcal{O}(\xi_2^3) \\
\Re \bigg(\frac{\eps'}{\eps} \bigg) &= 
\frac{\omega}{\sqrt{2} |\epsK|} (\xi_2 - \xi_0) \,.
\label{eq:e'/e:xi0}
\end{align}

There are two independent methods to determine $\xi_0$ in lattice QCD:
one is the indirect method and the other is the direct method.
In the indirect method, we determine $\xi_0$ using
Eq.~\eqref{eq:e'/e:xi0} with lattice QCD input $\xi_2$ and with
experimental results for $\eps'/\eps$, $\epsK$, and $\omega$.
In the direct method, we can determine $\xi_0$ directly using
lattice QCD results for $\Im A_0$ combined with experimental results
for $\Re A_0$.

Recently, RBC-UKQCD reported results for $\xi_2$ in Ref.~\cite{
  Blum:2015ywa}.
Using the indirect method, we can obtain the result for $\xi_0$
as in Table \ref{tab:xi0}.
Recently, RBC-UKQCD also reported results for $\Im A_0$ in Ref.~\cite{
  Bai:2015nea}.
Using the experimental value of $\Re A_0$, we can obtain $\xi_0$
directly from $\Im A_0$, which is summarized in Table \ref{tab:xi0}.
%
%
%
%
\begin{table}[htbp]
  \caption{ Input parameter $\xi_0$. }
  \label{tab:xi0}
  \renewcommand{\arraystretch}{1.2}
  \begin{ruledtabular}
    \begin{tabular}{lllc}
      parameter & method & value & Ref. \\ \hline
      $\xi_0$ & indirect & $-1.63(19) \times 10^{-4}$ & \cite{Blum:2015ywa} 
      \\
      $\xi_0$ & direct  & $-0.57(49) \times 10^{-4}$  & \cite{Bai:2015nea}
    \end{tabular}
  \end{ruledtabular}
\end{table}

\begin{figure}[htbp]
  \includegraphics[width=\linewidth]{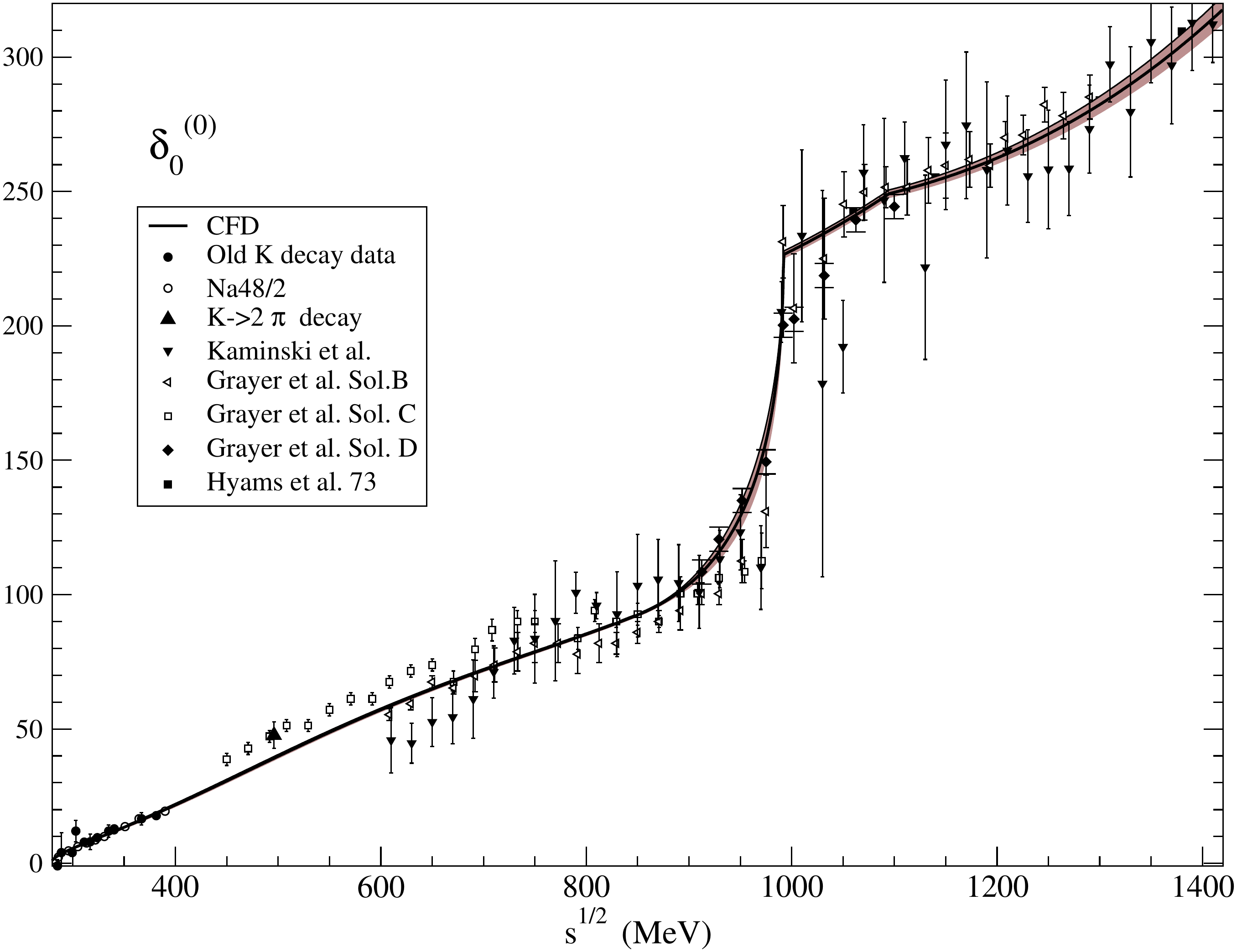}
  \caption{Experimental results for $\delta_0$. We borrow this plot from
  Ref.~\cite{GarciaMartin:2011cn}.}
  \label{fig:d0-exp}
\end{figure}

In Ref.~\cite{Bai:2015nea} RBC-UKQCD also reported the S-wave
$\pi-\pi$ scattering phase shift for the $I=0$ channel: $\delta_0 =
23.8(49)(12)$.
This value is $3.0\sigma$ lower than the conventional results for
$\delta_0$ in Refs.~\cite{GarciaMartin:2011cn} (KPY-2011) and
\cite{ Colangelo:2001df, Colangelo2016:MITP} (CGL-2001), and
\cite{ DescotesGenon:2001tn}.
The values for $\delta_0$ are summarized in Table \ref{tab:d0}.
%
%
%
%
\begin{table}[ht]
  \caption{Results for $\delta_0$}
  \label{tab:d0}
  \renewcommand{\arraystretch}{1.2}
  \begin{ruledtabular}
    \begin{tabular}{llr}
      Collaboration & $\delta_0$ & Ref. \\ \hline
      RBC-UKQCD-2016  & $23.8(49)(12){}^{\circ}$ & \cite{Bai:2015nea}
      \\
      KPY-2011 & 39.1(6)${}^{\circ}$ & \cite{GarciaMartin:2011cn}
      \\
      CGL-2001 & 39.2(15)${}^{\circ}$ & \cite{Colangelo:2001df,
        Colangelo2016:MITP}
    \end{tabular}
  \end{ruledtabular}
\end{table}
In Fig.~\ref{fig:d0-exp}, we show the experimental results for
$\delta_0$ with the fitting results of KPY-2011.
They (KPY-2011) used a singly subtracted Roy-like equation to do
the interpolation around $\sqrt{s} = m_K$ (the physical kaon mass).
Their fitting to the experimental data works well from the threshold to
$\sqrt{s} \cong 800 \MeV$.
In this range they use the singly subtracted Roy-like equation to do
the fitting.
%
%
%
%

\begin{figure}[htbp]
  \renewcommand{\subfigcapskip}{0.5em}
  \renewcommand{\subfiglabelskip}{0.3em}
  \subfigure[$\delta_0$]{
    \includegraphics[width=0.99\linewidth]{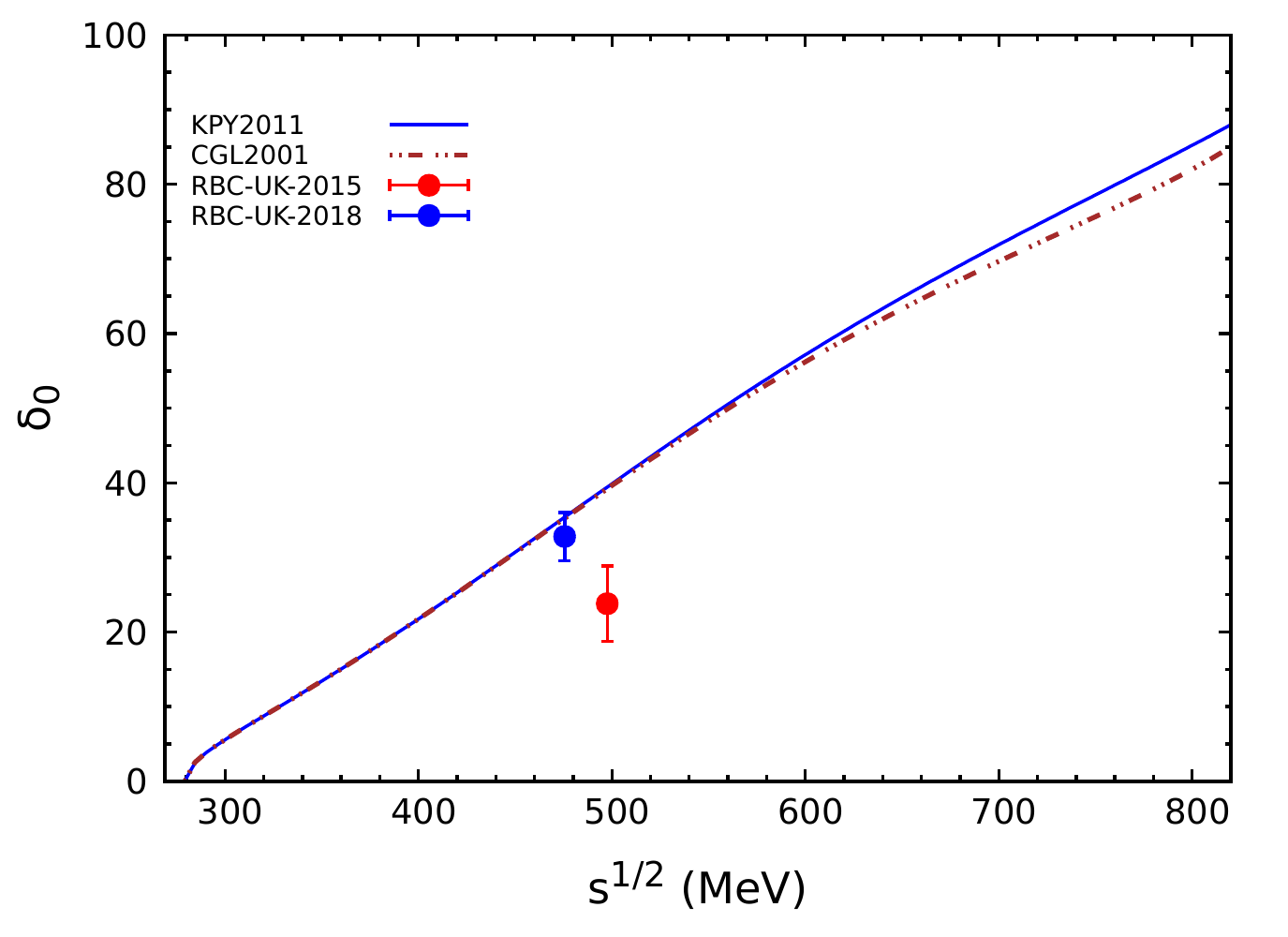}
    \label{fig:d0-rbc}
  }
  \subfigure[{$\delta_2$}]{
    \includegraphics[width=0.99\linewidth]{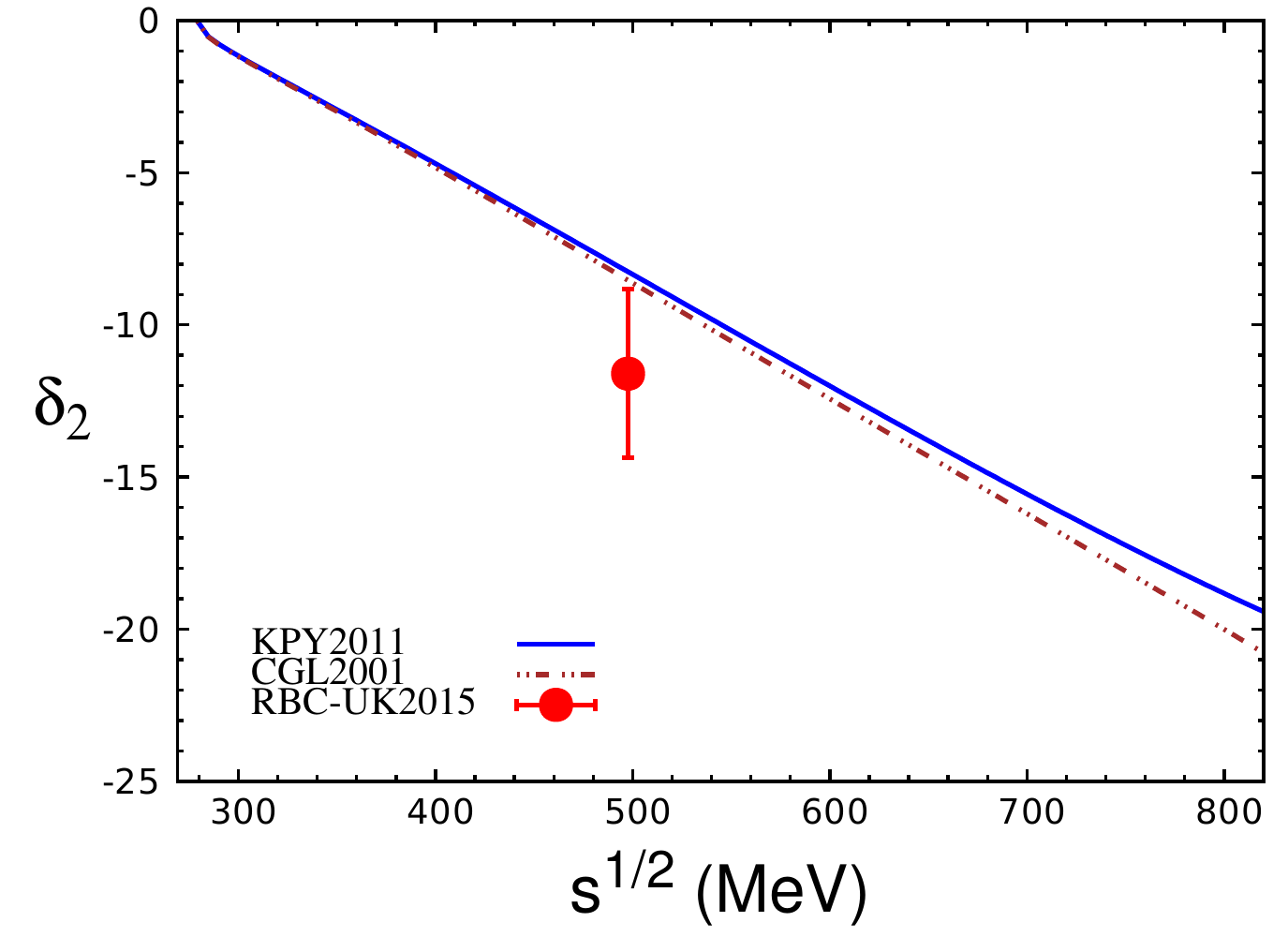}
    \label{fig:d2-rbc}
  }
  \caption{S-wave $\pi-\pi$ scattering phase shifts $\delta_I$ for
    \subref{fig:d0-rbc} $I=0$ and \subref{fig:d2-rbc} $I=2$ channels.}
  \label{fig:d0-d2}
\end{figure}

In Fig.~\ref{fig:d0-d2}, we show the fitting results of both KPY-2011
and CGL-2001 as well as the results of RBC-UKQCD.
There is essentially no difference between KPY-2011 and CGL-2001 in
the region near $\sqrt{s} = m_K$.
As one can see in Fig.~\ref{fig:d0-d2}\;\subref{fig:d0-rbc}, we
observe the $3.0\sigma$ tension for $\delta_0$ between RBC-UKQCD and
KPY-2011.
In the case of $\delta_2$, there is no difference between RBC-UKQCD
and KPY-2011 within statistical uncertainty as one can see in
Fig.~\ref{fig:d0-d2}\;\subref{fig:d2-rbc}.
Taking into account all the aspects, we conclude that the direct
calculation of $\Im A_0$ and $\xi_0$ by RBC-UKQCD in Ref.~\cite{
  Bai:2015nea} might have unresolved issues.
Indeed, preliminary results presented by RBC-UKQCD in Lattice 2018
suggests that this discrepancy might disappear with improved analysis
\cite{Wang:2018Latt}.
%
%
%
%

Therefore, we prefer the indirect method to the direct method for the
following two reasons.
The first reason is that the lattice QCD calculation of $\Im A_0$ is
much noisier than that of $\Im A_2$ thanks to many disconnected
diagrams.
The second reason is that the S-wave phase shift $\delta_0$ of the
$\pi-\pi$ scattering in Ref.~\cite{ Bai:2015nea} is lower by
$3.0\sigma$ than the conventional determination of $\delta_0$ in
Ref.~\cite{ Colangelo:2001df, GarciaMartin:2011cn,
  DescotesGenon:2001tn}, which indicates that there might be some
issues unresolved at present.
In Table \ref{tab:xi0}, we present results of $\xi_0$ determined
using both indirect and direct methods.
Here, we use the value of $\xi_0$ obtained using the indirect method.

One remaining caveat is that the $\xi_0$ and $\xi_2$ in Ref.~\cite{
  Blum:2015ywa} is calculated in the isospin symmetric limit.
The isospin breaking effects on $\eps'/\eps$ are studied in
Refs.~\cite{ Gardner:1999wb, Cirigliano:2003nn}.
These studies conclude that the isospin violation correction in the CP
violation correction for $\eps'$ is below 15\% within the
uncertainties of large $N_c$ estimates for the low energy constants.
Since $\xi_0$ has an effect of about $-7\%$ on $\epsK$, the isospin
violation effect maximum, 15\% of $\xi_0$ amounts to $\pm 1\%$
correction for $\epsK$.
Here, we neglect this effect completely without loss of generality
in our conclusion.

\subsection{$\xi_\text{LD}$}
\label{ssec:xi_LD}
The long distance (LD) effects on $\epsK$ are explained in
Sec.~\ref{ssec:LD}.
Hence, here we would like to summarize the recent progress in
calculating the LD effects in lattice QCD.
Lattice QCD tools to calculate the dispersive LD effect,
$\xi_\text{LD}$ are well established in Ref.~\cite{ Christ:2012se,
  Bai:2014cva, Christ:2015pwa}.
In addition, recently, there have been a number of attempts to
calculate $\xi_\text{LD}$ on the lattice \cite{ Christ:2015phf,
  Bai:2016gzv}.
In these attempts, RBC-UKQCD used pion mass of $329 \MeV$ and kaon
mass of $591 \MeV$.
Hence, the energy of the two pion state and three pion states are
heavier than the kaon mass.
Therefore, the sign of the denominator in Eq.~\eqref{eq:mLD} is
opposite to that of the physical contribution in which the two and
three pion state energy is lighter than the kaon mass.
Therefore, this attempt in Refs.~\cite{ Christ:2015phf, Bai:2016gzv}
belongs to the category of exploratory study rather than to that of
precision measurement.

The net contribution of $\xi_\text{LD}$ to $\epsK$ in
Eqs.~\eqref{eq:xiLD} and \eqref{eq:mLD} turns out to be of the same
order of magnitude as $\xi_0$ using chiral perturbation theory
\cite{Buras2010:PhysLettB.688.309}.
They claim that
\begin{align}
  \xi_\text{LD} &= \black{-0.4(3) \times \frac{\xi_0}{\sqrt{2}} }\,,
  \label{eq:xiLD:BGI}
\end{align}
where we use the indirect results of $\xi_0$ given in Table
\ref{tab:xi0}, including its error.
Here, we call this method the BGI estimate for $\xi_\text{LD}$.
This also indicates that $\xi_\text{LD}$ is at most a 4\% correction
to $\epsK$.
This claim is highly consistent with the estimate of about $2\%$ in
Ref.~\cite{ Christ:2012se, Christ:2014qwa}:
\begin{align}
  \xi_\text{LD} = (0 \pm 1.6)\%\,.
  \label{eq:xiLD:christ}
\end{align}
Here, we call this method the RBC-UKQCD estimate for $\xi_\text{LD}$.

In this paper, we use both of the above estimates of $\xi_\text{LD}$
with the BGI and RBC-UKQCD methods to determine $\epsK$.

\subsection{Top quark mass}
\label{ssec:top}
The pole mass of top quarks coming from Ref.~\cite{Patrignani:2016xqp}
is
\begin{align}
  M_t &= 173.5 \pm 1.1 \GeV
\end{align}
The pole and $\MSb$ masses are related as follows,
\begin{align}
  \frac{m_t(\mu)}{M_t} &= z(\mu) = \frac{ Z_\text{OS} }{ Z_{\MSb}}
\end{align}
where $m_t(\mu)$ is the $\MSb$ mass renormalized at scale $\mu$.
Here, $Z_\text{OS}$ is the renormalization factor in the on-shell
scheme, and $Z_{\MSb}$ is the renormalization factor in the $\MSb$
scheme.
The top scale-invariant quark mass $\mu_t$ is the $\MSb$ mass
$m_t(\mu)$ with the scale $\mu$ set equal to the scale-invariant mass,
\begin{align}
  \mu_t = m_t(\mu_t) &= 163.65 \pm 1.05 \pm 0.17 \GeV
  \label{eq:top-mass-SI}
\end{align}
where we use the four-loop perturbation formula for $z(\mu_t)$.
Details on the four-loop conversion formula are described in Appendix
\ref{app:top-mass}.
In Eq.~\eqref{eq:top-mass-SI}, the first error comes from the error of
the top pole mass, and the second error represents the uncertainty due
to truncation of higher loops in the conversion formula which is
estimated as the difference in $\mu_t$ between the 3-loop and 4-loop
formulas.
We have neglected the renormalon ambiguity and corrections due to the
three-loop fermion mass such as $m_b$ (bottom quark mass) and $m_c$
(charm quark mass).

\subsection{Other Input Parameters}
\label{ssec:other}
For the higher order QCD corrections $\eta_{cc}$, $\eta_{ct}$, and
$\eta_{tt}$, we use the same values as in Ref.~\cite{ Bailey:2015tba}.
They are summarized in Table \ref{tab:eta}.

\begin{table}[h]
  \caption{Higher order QCD corrections: $\eta_{cc}$, $\eta_{tt}$, and
    $\eta_{ct}$.}
  \label{tab:eta}
  \renewcommand{\arraystretch}{1.2}
  \begin{ruledtabular}
    \begin{tabular}[b]{@{\quad} c @{\qquad}l @{\qquad}c @{\quad}}
      Input & Value & Ref. \\ \hline
      $\eta_{cc}$ & $1.72(27)$   & \cite{Bailey:2015tba} 
      \\
      $\eta_{tt}$ & $0.5765(65)$ & \cite{Buras2008:PhysRevD.78.033005} 
      \\
      $\eta_{ct}$ & $0.496(47)$  & \cite{Brod2010:prd.82.094026}
    \end{tabular}
  \end{ruledtabular}
\end{table}

Other input parameters are summarized in Table \ref{tab:other}.
They are the same as Ref.~\cite{ Bailey:2015tba} except for charm
quark mass $m_c(m_c)$, the kaon mass $m_{K^0}$, and the kaon decay
constant $F_K$.
For the charm quark mass, we use the HPQCD results of $m_c(m_c)$
reported in Ref.~\cite{ Chakraborty:2014aca}.
For the kaon mass, we use the updated results of Particle Data Group
(PDG) in Ref.~\cite{Patrignani:2016xqp}.
For the kaon decay constant, we use the updated results of PDG
reported in Ref.~\cite{Patrignani:2016xqp}, which are obtained
from the FLAG data \cite{Aoki:2016frl}.

\begin{table}[h]
  \caption{Other input parameters.}
  \label{tab:other}
  \renewcommand{\arraystretch}{1.2}
  \begin{ruledtabular}
    \begin{tabular}{@{\quad} c @{\qquad} l @{\qquad} c @{\quad}}
      Input & Value & Ref. \\ \hline
      $G_{F}$
      & $1.1663787(6) \times 10^{-5}$ GeV$^{-2}$
      &\cite{Patrignani:2016xqp} \\ \hline
      $M_{W}$
      & $80.385(15)$ GeV
      &\cite{Patrignani:2016xqp} \\ \hline
      $m_{c}(m_{c})$
      & $1.2733(76)$ GeV
      &\cite{Chakraborty:2014aca} \\ \hline
      $\theta$
      & $43.52(5)^{\circ}$
      &\cite{Patrignani:2016xqp} \\ \hline
      $m_{K^{0}}$
      & $497.611(13)$ MeV
      &\cite{Patrignani:2016xqp} \\ \hline
      $\Delta M_{K}$
      & $3.484(6) \times 10^{-12}$ MeV
      &\cite{Patrignani:2016xqp} \\ \hline
      $F_K$
      & $155.6(4)$ MeV
      &\cite{Patrignani:2016xqp}
    \end{tabular}
  \end{ruledtabular}
\end{table}

\section{Results} 
\label{sec:result}

\subsection{RBC-UKQCD estimate for $\xi_\text{LD}$}
\label{ssec:nhc}

\begin{figure*}[htbp]
  \renewcommand{\subfigcapskip}{0.5em}
  \renewcommand{\subfiglabelskip}{0.3em}
  \subfigure[Exclusive $\Vcb$]{
    \includegraphics[width=0.47\linewidth]
                    {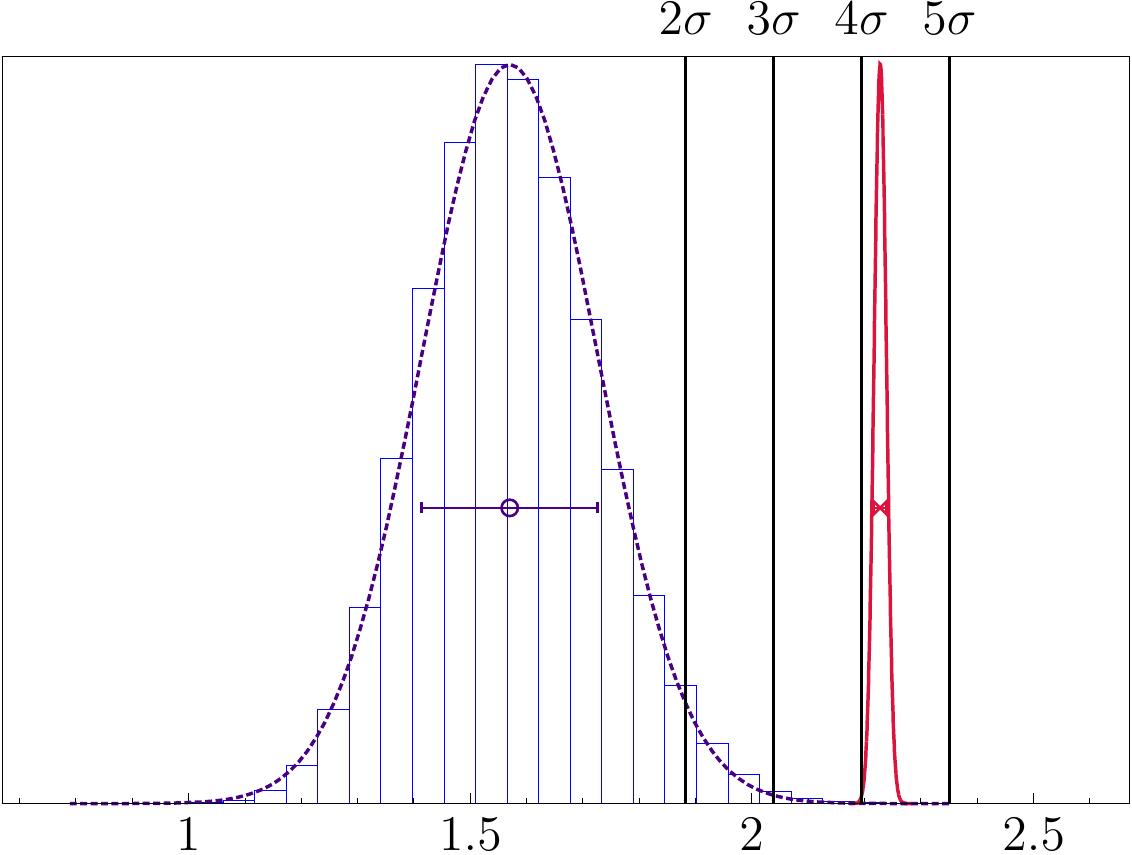}
    \label{fig:epsK-ex:nhc}
    }
  \hfill
  \subfigure[Inclusive $\Vcb$]{
    \includegraphics[width=0.47\linewidth]
                    {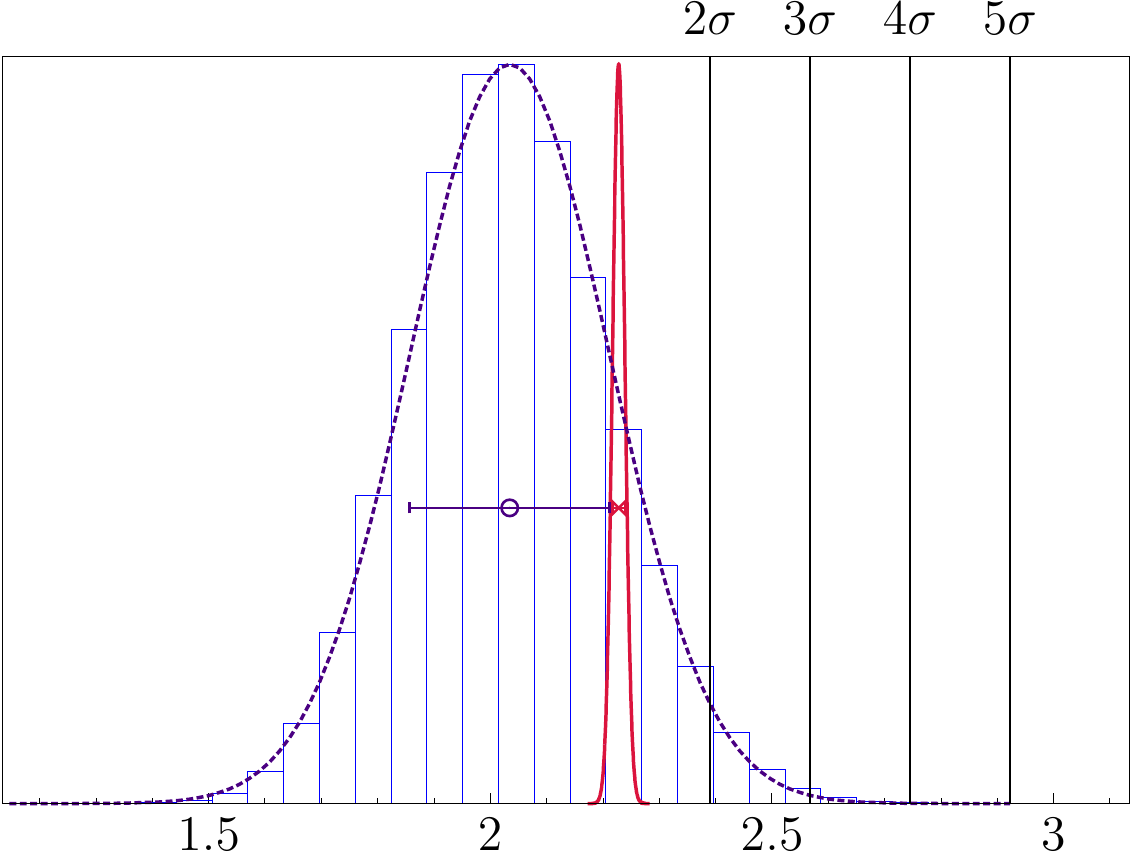}
    \label{fig:epsK-in:nhc}
    }
  \caption{$|\epsK|$ with \subref{fig:epsK-ex:nhc} exclusive $\Vcb$ (left)
    and \subref{fig:epsK-in:nhc} inclusive $\Vcb$ (right) in units of
    $1.0\times 10^{-3}$. Here, we use the FLAG-2017 $\BK$, AOF for the
    Wolfenstein parameters, and the RBC-UKQCD estimate in
    Eq.~\eqref{eq:xiLD:christ} for $\xi_\text{LD}$. The red curve
    represents the experimental results for $\epsK$ and the blue curve
    represents the theoretical results for $\epsK$ calculated directly
    from the standard model.}
  \label{fig:epsK:cmp:nhc}
\end{figure*}

In Fig.~\ref{fig:epsK:cmp:nhc}, we present results for $\epsK$ calculated
directly from the standard model with the lattice QCD inputs described
in Section \ref{sec:input}.
In Fig.~\ref{fig:epsK:cmp:nhc}\;\subref{fig:epsK-ex:nhc}, the blue
curve which encircles the histogram represents the theoretical
evaluation of $\epsK$ using the FLAG-2017 $\BK$, AOF for Wolfenstein
parameters, and exclusive $\Vcb$ which corresponds to ex-combined in
Table \ref{tab:Vcb}\;\subref{tab:ex-Vcb}, and the RBC-UKQCD estimate for
$\xi_\text{LD}$.
The red curve in Fig.~\ref{fig:epsK:cmp:nhc} represents the experimental
result for $\epsK$.
In Fig.~\ref{fig:epsK:cmp:nhc}\;\subref{fig:epsK-in:nhc}, the blue
curve represents the same as in
Fig.~\ref{fig:epsK:cmp:nhc}\;\subref{fig:epsK-ex:nhc} except for using
the inclusive $\Vcb$ which corresponds to 1S scheme in Table
\ref{tab:Vcb}\;\subref{tab:in-Vcb}.

The updated results for $|\epsK|$ are, in units of $1.0\times 10^{-3}$,
presented in Table \ref{tab:epsK:nhc}.
From Table \ref{tab:epsK:nhc}, we observe that the theoretical
evaluation of $|\epsK|$ with lattice QCD inputs (with exclusive
$\Vcb$), which corresponds to $|\epsK|^\text{SM}_\text{excl}$, has
$4.2\sigma$ tension with the experimental result $|\epsK|^\text{Exp}$,
while there is no tension in the inclusive $\Vcb$ channel (heavy quark
expansion based on the OPE and QCD sum rules).
%
%
%
\begin{table}[htbp]
  \caption{$|\epsK|$ values in units of $1.0 \times 10^{-3}$. The
    superscript SM represents the standard model. The subscript excl
    (incl) represents exclusive (inclusive) $\Vcb$. The superscript Exp
    represents the experimental result. We use the same input
    parameters as in Fig.~\ref{fig:epsK:cmp:nhc}.  }
  \label{tab:epsK:nhc}
  \renewcommand{\arraystretch}{1.4}
  \begin{ruledtabular}
    \begin{tabular}{lll}
      parameter & method & value \\ \hline
      $|\epsK|^\text{SM}_\text{excl}$ & exclusive $\Vcb$ & $1.570 \pm 0.156$ 
      \\
      $|\epsK|^\text{SM}_\text{incl}$ & inclusive $\Vcb$ & $2.035 \pm 0.178$ 
      \\
      $|\epsK|^\text{Exp}$            & experiment       & $2.228 \pm 0.011$
    \end{tabular}
  \end{ruledtabular}
\end{table}

\begin{figure*}[htbp]
  \renewcommand{\subfigcapskip}{0.5em}
  \renewcommand{\subfiglabelskip}{0.3em}
  \subfigure[Time evolution of $\Delta \epsK/\sigma$]{
    \includegraphics[width=0.475\linewidth]
                    {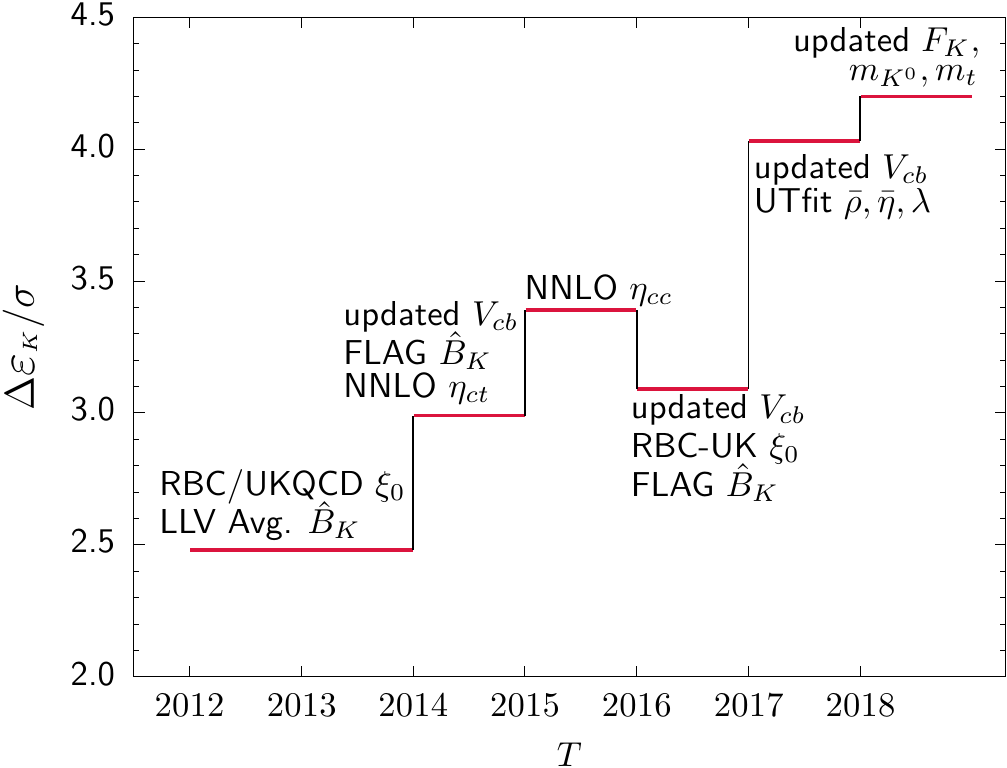}
    \label{fig:depsK:nhc:his}
    }
  \hfill
  \subfigure[Time evolution of the average and error of $\Delta\epsK$]{
    \includegraphics[width=0.45\linewidth]
                    {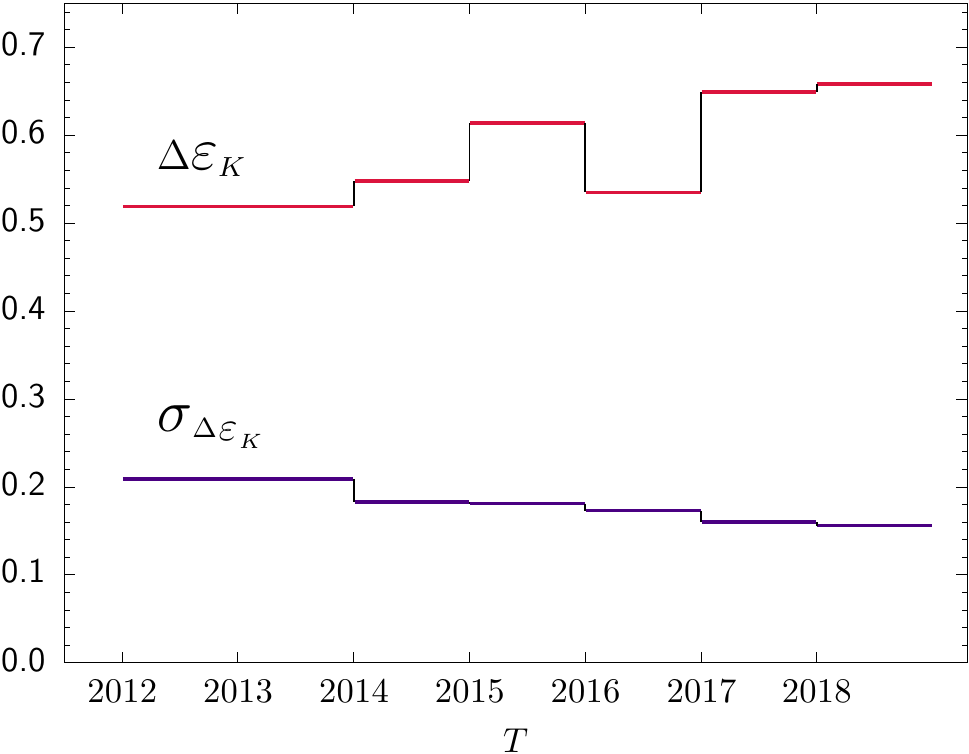}
    \label{fig:depsK+sigma:nhc:his}
    }
  \caption{ Time history of \subref{fig:depsK:nhc:his}
    $\Delta\epsK/\sigma$, and \subref{fig:depsK+sigma:nhc:his}
    $\Delta\epsK$ and $\sigma_{\Delta\epsK}$.  We define $\Delta \epsK
    \equiv |\epsK|^\text{Exp} - |\epsK|^\text{SM}_\text{excl}$.  Here,
    $\sigma = \sigma_{\Delta\epsK}$ represents the error of
    $\Delta\epsK$.  The $\Delta\epsK$ is obtained using the same input
    parameters as in
    Fig.~\ref{fig:epsK:cmp:nhc}\;\subref{fig:epsK-ex:nhc} for the
    exclusive $\Vcb$ channel. }
  \label{fig:depsK:sum:nhc:his}
\end{figure*}

In Fig.~\ref{fig:depsK:sum:nhc:his}\;\subref{fig:depsK:nhc:his}, we
plot the $\Delta\epsK \equiv |\epsK|^\text{Exp} -
|\epsK|^\text{SM}_\text{excl}$ in units of $\sigma$ (which is the total
error of $\Delta\epsK$) as the time evolves starting from 2012.
We began to monitor $\Delta\epsK$ in 2012 when several lattice QCD
results for $\BK$ obtained using different discretization methods
for the light and strange quarks became consistent with one another
within one sigma.
In 2012, $\Delta\epsK$ was $2.5\sigma$, but now it is $4.2\sigma$.
To understand the change of $\Delta\epsK/\sigma$ with respect to time,
we have performed an additional analysis on the average and error.

In Fig.~\ref{fig:depsK:sum:nhc:his}\;\subref{fig:depsK+sigma:nhc:his},
we plot the time evolution of the average $\Delta\epsK$ and the
error $\sigma_{\Delta\epsK}$.
Here, we find that the average of $\Delta\epsK$ has increased with
some fluctuations by 27\% during the period of
2012--2018, and its error $\sigma_{\Delta\epsK}$ has decreased
monotonically by 25\% in the same period.
These two effects interfere constructively to produce the $4.2\sigma$
tension in $\Delta\epsK$ in 2018.
We can understand the monotonic decrease in $\sigma_{\Delta\epsK}$
in the following way.
As time goes on, the lattice QCD calculations are becoming more
precise and the experimental results also are becoming more accurate,
which constructively leads to the monotonic decrease in
$\sigma_{\Delta\epsK}$.

In Table \ref{tab:err-budget:nhc}, we present the error budget for
$|\epsK|^\text{SM}_\text{excl}$.
Here, we find that the largest error in $|\epsK|^\text{SM}_\text{excl}$
comes from $\Vcb$, while the errors coming from $\bar{\eta}$
and $\eta_{ct}$ are sub-dominant.
Hence, if we are to see a gap $\Delta\epsK$ greater than $5.0\sigma$,
it is essential to reduce the error in $\Vcb$ significantly.
%

\begin{table}[tb!]
  \caption{ Error budget for $|\epsK|^\text{SM}_\text{excl}$ obtained
    using the AOF method for the Wolfenstein parameters, the exclusive
    $\Vcb$, the FLAG-2017 $\BK$, and the RBC-UKQCD estimate for
    $\xi_\text{LD}$. Here, the values are fractional contributions to
    the total error obtained using the formula in Ref.~\cite{
      Bailey:2015tba}. }
  \label{tab:err-budget:nhc}
  \renewcommand{\arraystretch}{1.2}
  \begin{ruledtabular}
  \begin{tabular}{ccc}
    source          & error (\%)  & memo \\
    \hline
    $\Vcb$          & 31.4        & ex-combined \\
    $\bar{\eta}$    & 26.8        & AOF \\
    $\eta_{ct}$     & 21.5        & $c-t$ Box \\
    $\eta_{cc}$     &  9.1        & $c-c$ Box \\
    $\bar{\rho}$    &  4.0        & AOF \\
    $\xi_\text{LD}$ &  2.5        & RBC/UKQCD \\
    $\hat{B}_K$     &  1.9        & FLAG \\
    $\eta_{tt}$     &  0.77       & $t-t$ Box \\
    $\xi_0$         &  0.70       & RBC/UKQCD \\
    $m_t$           &  0.67       & $m_t(m_t)$ \\
    $\lambda$       &  0.33       & $\Vus$ \\
    $\vdots$        & $\vdots$    & $\vdots$
  \end{tabular}
  \end{ruledtabular}
\end{table}

In exclusive $\Vcb$, there are two major error sources: one is
experimental and the other is theoretical.
The experimental error is discussed in Section \ref{ssec:Vcb}, and
the resolution is beyond the scope of this paper.
The largest error in the theoretical part of $\Vcb$ comes from
the heavy quark discretization error (HQDE) for the charm quark
in lattice QCD.
If one use the Fermilab action, the HQDE is about 1.0\%, which is
significantly larger than any other error in the theoretical side.
In order to reduce the HQDE by a factor of $\sim 1/5$, there are
on-going efforts to use the OK action to calculate the $B \to D^{(*)}$
semileptonic form factors in Refs.~\cite{ Bailey:2017xjk,
  Bailey:2017zgt, Jeong:2016eot}.

\subsection{BGI estimate for $\xi_\text{LD}$}
\label{ssec:bgi}
Here, we present the results obtained using the BGI estimate
in Eq.~\eqref{eq:xiLD:BGI} for $\xi_\text{LD}$.

\begin{figure*}[htbp]
\renewcommand{\subfigcapskip}{0.5em}
\renewcommand{\subfiglabelskip}{0.3em}
  \subfigure[Exclusive $\Vcb$]{
    \includegraphics[width=0.47\linewidth]
                    {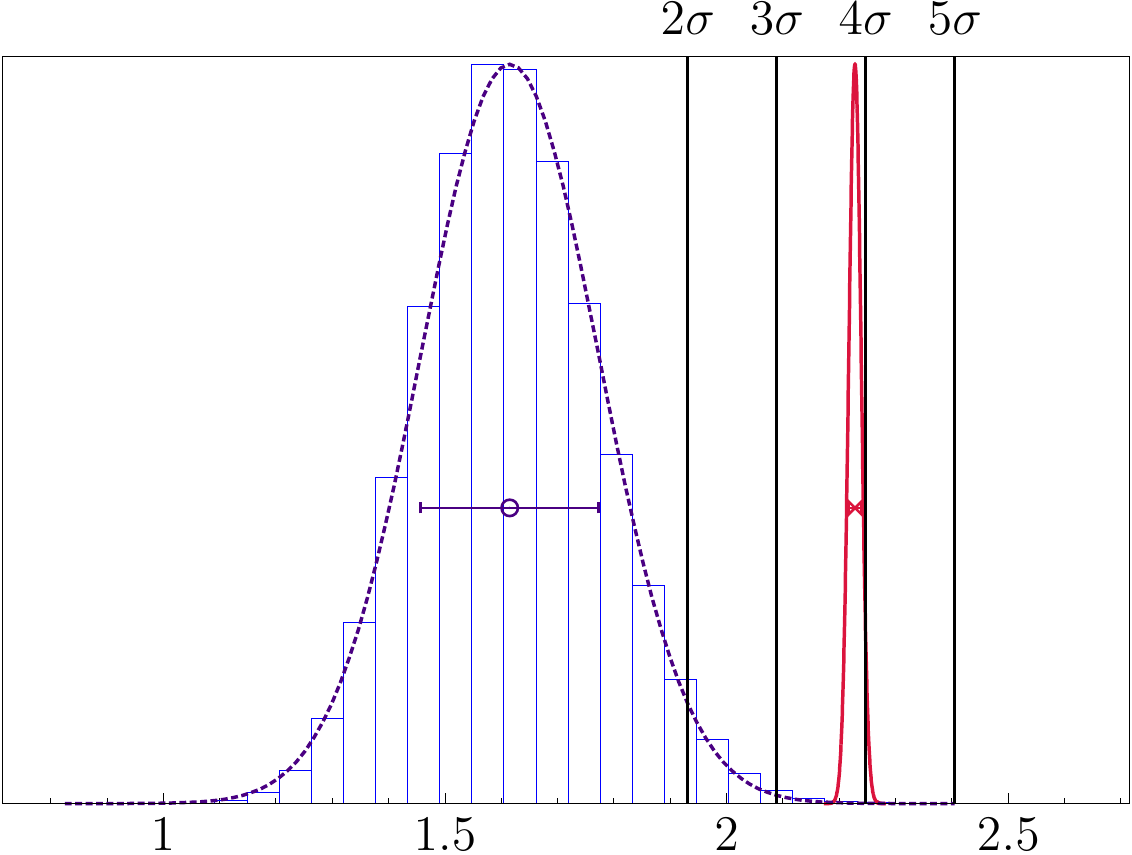}
    \label{fig:epsK-ex:bgi}
    }
  \hfill
  \subfigure[Inclusive $\Vcb$]{
    \includegraphics[width=0.47\linewidth]
                    {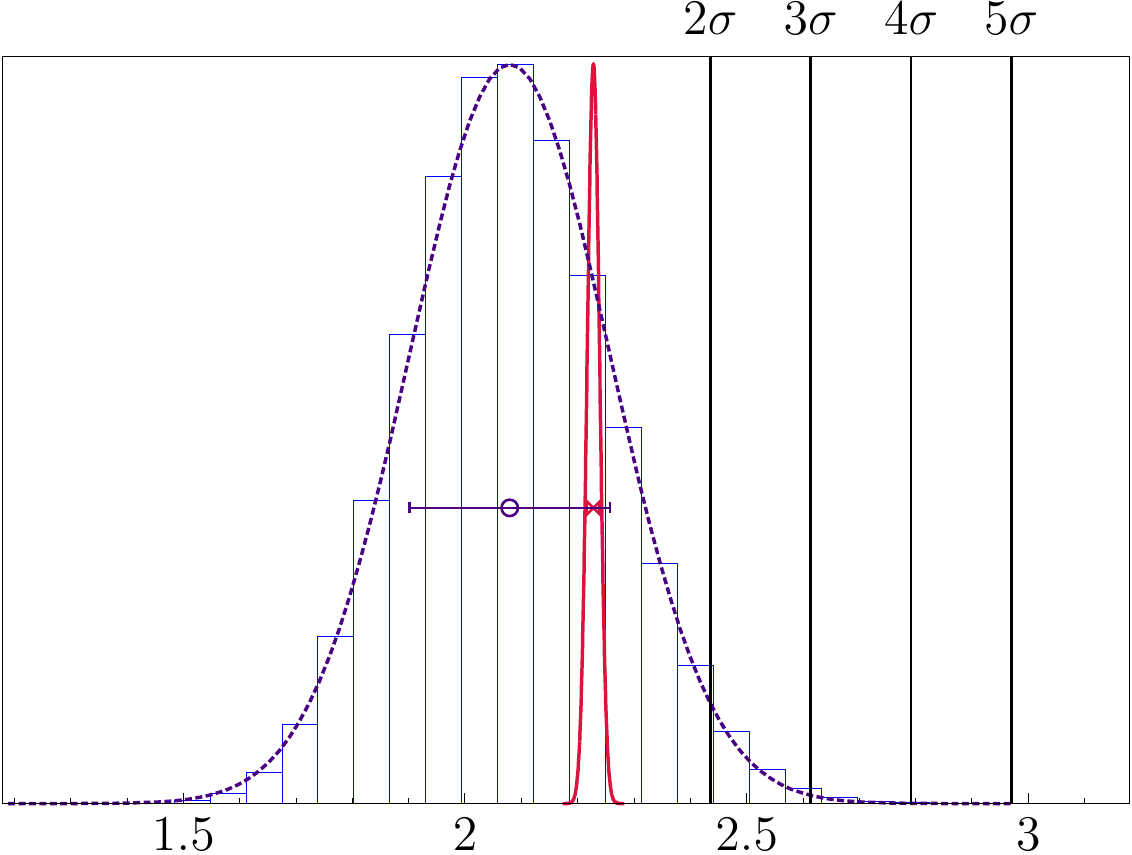}
    \label{fig:epsK-in:bgi}
    }
  \caption{$|\epsK|$ with \subref{fig:epsK-ex:bgi} exclusive $\Vcb$
    (left) and \subref{fig:epsK-in:bgi} inclusive $\Vcb$ (right) in
    units of $1.0\times 10^{-3}$. Here, we use the FLAG-2017 $\BK$,
    AOF for the Wolfenstein parameters, and the BGI estimate in
    Eq.~\eqref{eq:xiLD:BGI} for $\xi_\text{LD}$. The red curve
    represents the experimental results for $\epsK$ and the blue curve
    represents the theoretical results for $\epsK$ calculated directly
    from the standard model.}
  \label{fig:epsK:cmp:bgi}
\end{figure*}

In Fig.~\ref{fig:epsK:cmp:bgi}\;\subref{fig:epsK-ex:bgi}, the blue
curve represents the theoretical evaluation of $|\epsK|$ directly from
the standard model (SM) using the same input parameters as in
Fig.~\ref{fig:epsK:cmp:nhc}\;\subref{fig:epsK-ex:nhc} except for the
BGI estimate in Eq.~\eqref{eq:xiLD:BGI} for $\xi_\text{LD}$.
The red curve in Fig.~\ref{fig:epsK:cmp:bgi} represents the experimental
result for $|\epsK|$.
In Fig.~\ref{fig:epsK:cmp:bgi}\;\subref{fig:epsK-in:bgi}, the blue
curve represents the same as in
Fig.~\ref{fig:epsK:cmp:bgi}\;\subref{fig:epsK-ex:bgi} except for using
inclusive $\Vcb$ (1S scheme in Table
\ref{tab:Vcb}\;\subref{tab:in-Vcb}).

Results for $|\epsK|$ in Fig.~\ref{fig:epsK:cmp:bgi} are summarized in
Table \ref{tab:epsK:bgi}.
From Table \ref{tab:epsK:bgi}, we find that the value for
$|\epsK|^\text{SM}_\text{excl}$ (the theoretical evaluation of $|\epsK|$
with lattice QCD inputs such as exclusive $\Vcb$) has $3.9\sigma$
tension with the experimental result $|\epsK|^\text{Exp}$, whereas
there is no tension in the inclusive $\Vcb$ channel (with heavy quark
expansion and QCD sum rules).
%
%
%
%
\begin{table}[htbp]
  \caption{$|\epsK|$ values in units of $1.0 \times 10^{-3}$. The
    superscripts and subscripts follow the same notation as in Table
    \ref{tab:epsK:nhc}. We use the same input parameters as in
    Fig.~\ref{fig:epsK:cmp:bgi} to determine $|\epsK|$.}
  \label{tab:epsK:bgi}
  \renewcommand{\arraystretch}{1.4}
  \begin{ruledtabular}
    \begin{tabular}{lll}
      parameter & method & value \\ \hline
      $|\epsK|^\text{SM}_\text{excl}$ & exclusive $\Vcb$ & $1.615 \pm 0.158$ 
      \\
      $|\epsK|^\text{SM}_\text{incl}$ & inclusive $\Vcb$ & $2.079 \pm 0.178$ 
      \\
      $|\epsK|^\text{Exp}$            & experiment       & $2.228 \pm 0.011$
    \end{tabular}
  \end{ruledtabular}
\end{table}

\begin{figure*}[htbp]
  \renewcommand{\subfigcapskip}{0.5em}
  \renewcommand{\subfiglabelskip}{0.3em}
  \subfigure[Time evolution of $\Delta \epsK/\sigma$]{
    \includegraphics[width=0.475\linewidth]
                    {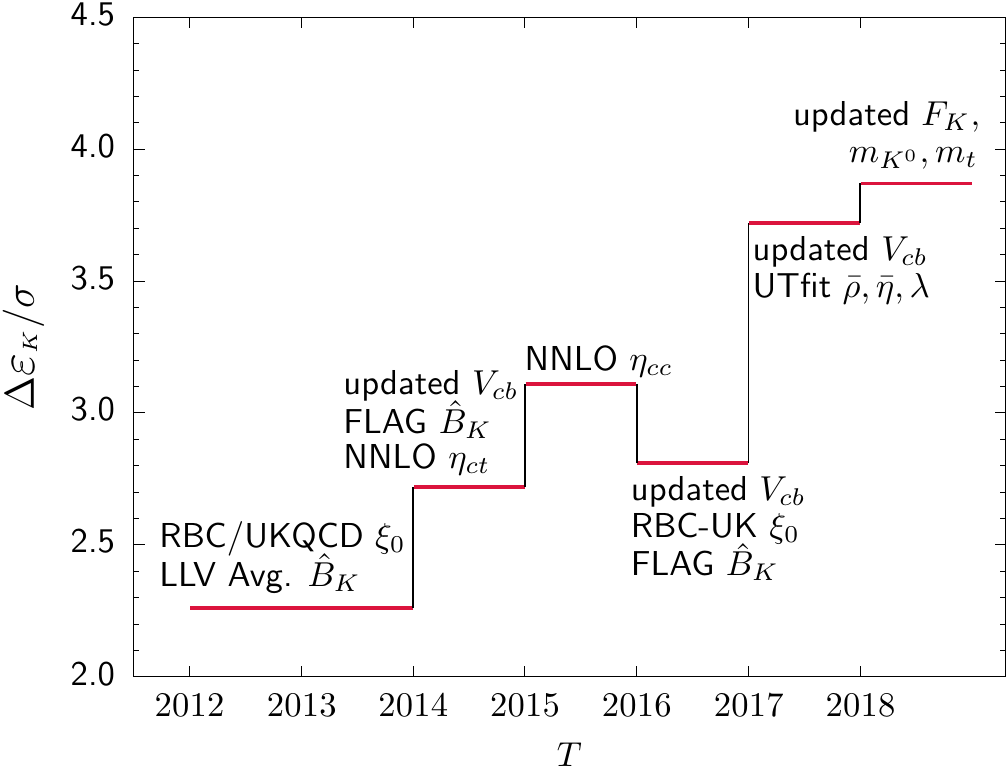}
    \label{fig:depsK:bgi:his}
    }
  \hfill
  \subfigure[Time evolution of  the average and error of $\Delta\epsK$]{
    \includegraphics[width=0.45\linewidth]
                    {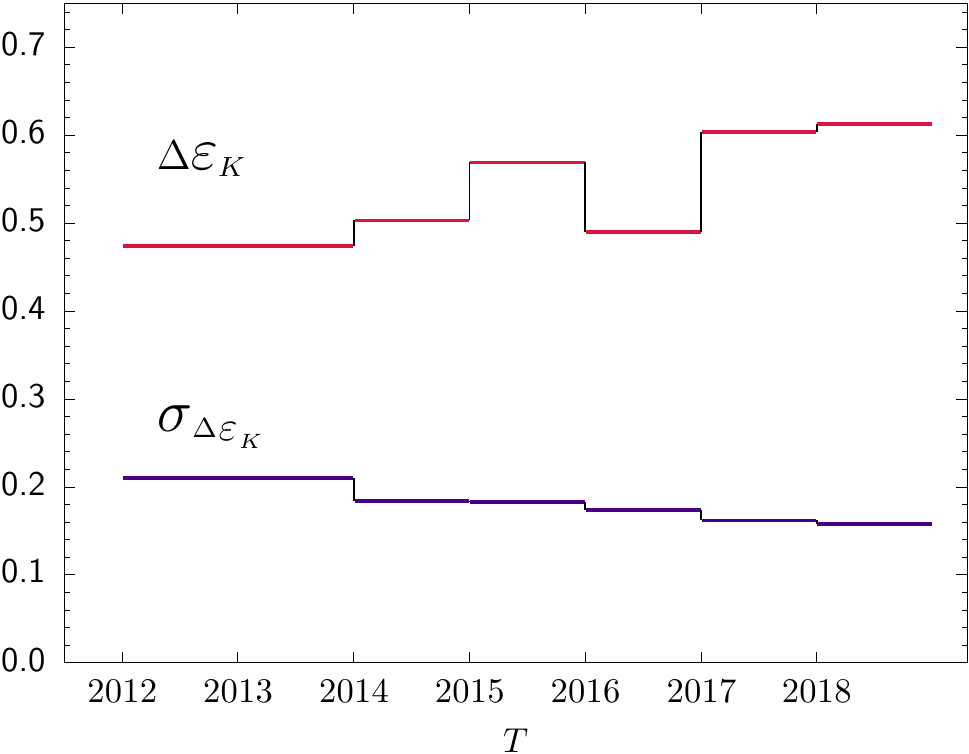}
    \label{fig:depsK+sigma:bgi:his}
    }
  \caption{ Time history of \subref{fig:depsK:bgi:his}
    $\Delta\epsK/\sigma$, and \subref{fig:depsK+sigma:bgi:his}
    $\Delta\epsK$ and $\sigma_{\Delta\epsK}$.  We define $\Delta \epsK
    \equiv |\epsK|^\text{Exp} - |\epsK|^\text{SM}_\text{excl}$.  Here,
    $\sigma = \sigma_{\Delta\epsK}$ represents the error of
    $\Delta\epsK$.  The $\Delta\epsK$ is obtained using the same input
    parameters as in
    Fig.~\ref{fig:epsK:cmp:bgi}\;\subref{fig:epsK-ex:bgi} for the
    exclusive $\Vcb$ channel. }
  \label{fig:depsK:sum:bgi:his}
\end{figure*}

In Fig.~\ref{fig:depsK:sum:bgi:his}\;\subref{fig:depsK:bgi:his}, we
plot $\Delta\epsK \equiv |\epsK|^\text{Exp} -
|\epsK|^\text{SM}_\text{excl}$ in units of $\sigma$ (the total error of
$\Delta\epsK$) as a function of time starting from 2012.
In 2012, $\Delta\epsK$ was $2.3\sigma$, but now it is $3.9\sigma$.
To understand this transition, we have done an additional analysis
on the average and error.

In Fig.~\ref{fig:depsK:sum:bgi:his}\;\subref{fig:depsK+sigma:bgi:his},
we plot the time evolution of the average and error for $\Delta\epsK$.
Here, we find that the average of $\Delta\epsK$ has increased by 29\%
with some fluctuations during the period of
2012--2018, and its error has decreased by 25\% monotonically in the
same period.
These two effect has produced, constructively, the $3.9\sigma$ tension
in $\Delta\epsK$.

In Table \ref{tab:err-budget:bgi}, we present the error budget for
$|\epsK|^\text{SM}_\text{excl}$.
Here, we find that the largest error in $|\epsK|^\text{SM}_\text{excl}$
still comes from $\Vcb$.
Here, note that the error from $\xi_\text{LD}$ (the BGI estimate)
is larger than that from $\bar{\rho}$, which is different from Table
\ref{tab:err-budget:nhc}.
In summary, if we are to observe the gap $\Delta\epsK$ greater than
$5.0\sigma$, it is essential to reduce the error in $\Vcb$
significantly.
%

\begin{table}[tb!]
  \caption{ Error budget for $|\epsK|^\text{SM}_\text{excl}$ obtained
    using the AOF method for the Wolfenstein parameters, the exclusive
    $\Vcb$, the FLAG-2017 $\BK$, and the BGI estimate for
    $\xi_\text{LD}$. Here, the values are fractional contributions to
    the total error obtained using the formula in Ref.~\cite{
      Bailey:2015tba}.  }
  \label{tab:err-budget:bgi}
  \renewcommand{\arraystretch}{1.2}
  \begin{ruledtabular}
  \begin{tabular}{ccc}
    source          & error (\%)  & memo \\
    \hline
    $\Vcb$          & 30.7        & ex-combined \\
    $\bar{\eta}$    & 26.2        & AOF \\
    $\eta_{ct}$     & 21.0        & $c-t$ Box \\
    $\eta_{cc}$     &  8.9        & $c-c$ Box \\
    $\xi_\text{LD}$ &  4.7        & BGI estimate \\
    $\bar{\rho}$    &  3.9        & AOF \\
    $\hat{B}_K$     &  1.8        & FLAG \\
    $\eta_{tt}$     &  0.76       & $t-t$ Box \\
    $\xi_0$         &  0.69       & RBC/UKQCD \\
    $m_t$           &  0.66       & $m_t(m_t)$ \\
    $\lambda$       &  0.32       & $\Vus$ \\
    $\vdots$        & $\vdots$    & $\vdots$
  \end{tabular}
  \end{ruledtabular}
\end{table}
%
%
%

%

\section{Conclusion}
\label{sec:conclude}
In this paper, we find that there exists a remarkable gap of $4.2\sigma
\sim 3.9\sigma$ in $\epsK$ between experiment and the SM theory with
lattice QCD inputs.
The upper bound of $4.2\sigma$ tension is obtained with the RBC-UKQCD
estimate for $\xi_\text{LD}$.
The lower bound of the $3.9\sigma$ tension is obtained when we use the
BGI estimate for $\xi_\text{LD}$.
In the BGI estimate \cite{Buras2010:PhysLettB.688.309}, they added
50\% more error to be on the safe side and more conservative.
Even if we remove this 50\% bubble in the error of the BGI estimate,
we end up with the same tension of $3.9\sigma$.
To obtain this result, we choose the angle-only-fit (AOF), exclusive
$\Vcb$ from lattice QCD, and FLAG $\BK$ ($N_f=2+1$) from lattice QCD,
to determine the theoretical value for $\epsK$ directly from the SM.
In 2015, we reported a $3.4\sigma$ tension between
$|\eps_K|^\text{SM}_\text{excl}$ and $|\eps_K|^\text{Exp}$ \cite{
  Bailey:2015tba}, and the tension is $4.2\sigma$ at
present.\footnote{In 2015, we used the RBC-UKQCD estimate for
  $\xi_\text{LD}$.}
We find that the tension between $|\eps_K|^\text{SM}_\text{excl}$ and
$|\eps_K|^\text{Exp}$ continues to increase during the period of
2012--2018.
Part of the reason is that the uncertainties of results for the SM
input parameters continues to decrease monotonically.

\begin{table}[h]
  \renewcommand{\arraystretch}{1.2}
  \caption{Results for $\Delta\epsK$.}
  \label{tab:DepsK}
  \begin{ruledtabular}
    \begin{tabular}{l l l}
      year & Inclusive $\Vcb$ & Exclusive $\Vcb$ \\ \hline
      2015 & $0.33\sigma$     & $3.4\sigma$ \\
      2018 & $1.1\sigma$      & $4.2\sigma$
    \end{tabular}
  \end{ruledtabular}
\end{table}

In Table \ref{tab:DepsK}, we present how the values of $\Delta \epsK$
have changed from 2015 to 2018.
Here, we find that the positive shift of $\Delta \epsK$ is about the
same for the inclusive and exclusive values of $\Vcb$.
This reflects the changes of other input parameters since 2015.
We also note that there is no significant tension observed
yet for inclusive $\Vcb$, which is obtained using the heavy quark
expansion based on the QCD sum rules.

There has been an interesting claim \cite{ Bigi:2017njr,
  Grinstein:2017nlq} which has potential to resolve the issue of the
inconsistency between the exclusive and inclusive $\Vcb$.
However, this claim is far away from conclusive yet since it is based
on an analysis over a preliminary and specific subset of experimental
data.
We find that it would be highly desirable if an experimental group
were to perform a comprehensive reanalysis over the entire set of
experimental data for the $\bar{B} \to D^* \ell \bar{\nu}$ decays
using an alternative parametrization method for the form factors, and
compare results with those of CLN.
%

\begin{acknowledgments}
We would like to express our sincere gratitude to Carleton Detar, Aida
El-Khadra, and Andreas Kronfeld for helpful discussion.
We also would like to express sincere gratitude to Guido Martinelli
for providing to us the most updated results of UTfit.
The research of W.~Lee is supported by the Creative Research
Initiatives Program (No.~2017013332) of the NRF grant funded by the
Korean government (MEST).
W.~Lee would like to acknowledge the support from the KISTI
supercomputing center through the strategic support program for the
supercomputing application research [No.~KSC-2015-G2-0002].
Computations were carried out on the DAVID GPU clusters at Seoul
National University.
J.A.B. is supported by the Basic Science Research Program of the
National Research Foundation of Korea (NRF) funded by the Ministry of
Education (No.~2014027937).
\end{acknowledgments}

\appendix

\section{Brief summary on CLN and BGL}
\label{app:cln-bgl}
%

Let us consider $\bar{B} \to D^* \ell \bar{\nu}$ decays.
The recoil variable $w$ is defined as $w \equiv v_{B} \cdot v_{D^*}$,
where $v_{B}$ is the four velocity of the mother particle ($\bar{B}$
meson) and $v_{D^*}$ is that of the daughter particle ($D^*$ meson).
The differential decay rate \cite{Neubert:1993mb} is given by
\begin{align}
  \frac{d\Gamma(\bar{B} \to D^* \ell \bar{\nu})}{dw}
  = & \frac{G_F^2 m_{D^*}^3}{48\pi^3} (m_B - m_{D^*})^2
  \CL
  & \times \chi(w) \eta^2_{\text{EW}} \mathcal{F}^2(w) \Vcb^2
\end{align}
where $G_F$ is Fermi's constant, $\eta_{\text{EW}}$ is a small
electroweak correction, and $\mathcal{F}(w)$ is the form factor.
The kinematic factor $\chi(w)$ is
\begin{align}
  \chi(w) &= \sqrt{w^2-1} (w+1)^2 \times Y(w) \\
  Y(w) &= \bigg[ 1 + \frac{4w}{w+1}
    \frac{1 - 2 w r + r^2}{(1 - r)^2} \bigg]
\end{align}
where $r \equiv m_{D^*}/m_B$.
So far the formalism is quite general.
We may express the form factor as follows , without loss of
generality,
\begin{align}
  \mathcal{F}^2(w) =& h_{A_1}^2(w) \times \frac{1}{Y(w)} \times
  \CL &
  \bigg\{ 2 \frac{1 - 2 w r + r^2}{(1 - r)^2} \bigg[ 1 +
    \frac{w-1}{w+1}R_1^2(w)\bigg] +
  \CL &
  \bigg[ 1 + \frac{w-1}{1-r} \Big( 1-R_2(w) \Big) \bigg]^2 \bigg\}
  \label{eq:cln-ff-def}
\end{align}
In the CLN method \cite{Caprini:1997mu}, the form factor functions are
parametrized as follows,
\begin{align}
  h_{A_1}(w) =& h_{A_1}(1) \Big[1 - 8 \rho^2 z + (53\rho^2 -15) z^2
    \CL
    &- (231 \rho^2 - 91) z^3\Big]
  \\
  R_1(w) =& R_1(1) - 0.12 (w-1) + 0.05 (w-1)^2
  \\
  R_2(w) =& R_2(1) + 0.11 (w-1) - 0.06 (w-1)^2
\end{align}
where $z$ is a typical conformal mapping variable defined as
\begin{align}
  z = \frac{\sqrt{w+1}-\sqrt{2}}{\sqrt{w+1}+\sqrt{2}}
\end{align}
The basic idea of CLN is a zero-recoil expansion around $w=1$.
The slope and curvature of $R_1(w)$ and $R_2(w)$ are determined by
perturbation theory and $\mathcal{O}(1/M)$ corrections at the leading
order using Heavy Quark Effective Theory (HQET) \cite{
Neubert:1993mb}.
The original claim of CLN \cite{ Caprini:1997mu} is that the accuracy
of $h_{A_1}(w)$ is better than 2\%, which makes us become apprehensive
as we get $\Vcb$ in the precision level below 2\%.
In addition, the slope and curvature of $R_1(w)$ and $R_2(w)$ contain
truncation errors coming from $\mathcal{O}(\Lambda^2/m_c^2)$ and
$\mathcal{O}(\alpha_s \Lambda/m_c)$ corrections as well as
those uncertainties due to the QCD sum rules on which it is based
\cite{Bigi:2017njr}. 
Typically, $\Lambda/m_c \approx 1/3$ and $\alpha_s \approx 0.3$, which
implies that the accuracy of the slope and curvature in the ratio
$R_i(w)$ is only in the 10\% level.

Using the CLN method, experimentalists perform a four parameter fit of
$\eta_{\text{EW}} \mathcal{F}(1) \Vcb$, $\rho^2$, $R_1(1)$ and
$R_2(1)$ to some unfolded data in experiment \cite{Amhis:2016xyh}.
Since $\eta_{\text{EW}}$ is very well known and lattice QCD can
determine $\mathcal{F}(1)$ very precisely, we can determine $\Vcb$
from the experimental fits.
%

Let us switch the gear to BGL.
In the case of CLN, it is built on the basis of HQET and its perturbative
expansion.
Unlike CLN, BGL is an HQET-independent approach to the form factor
parametrization.
The basic idea of BGL is composed of three building blocks: dispersion
relationship, analytic continuation, and crossing symmetry.

Let us begin with the first building block: dispersion relation.
In QCD, consider the two point function of flavor changing current
$J_\mu = V_\mu$, $A_\mu$, or $(V-A)_\mu$, where $V_\mu = \bar{c}
\gamma_\mu b$ and $A_\mu = \bar{c} \gamma_\mu\gamma_5 b$.
\begin{align}
  \Pi^{\mu\nu}_J (q) &=
  (q^\mu q^\nu - q^2 g^{\mu\nu}) \Pi^T_J (q^2)
  + g^{\mu\nu} \Pi^L_J (q^2)
  \CL
  &\equiv i \int d^4x e^{i q \cdot x}
  \langle 0 | T J^\mu(x) [J^\nu(0)]^\dagger | 0 \rangle
\end{align}
In general, $\Pi^{T,L}_J(q^2)$ is not finite.
Hence, in order to obtain finite dispersion relations, we need
to make one or two subtractions as follows,
\begin{align}
  \chi^L_J(q^2) &= \frac{\partial \Pi^L_J}{\partial q^2}
  = \frac{1}{\pi} \int_0^\infty dt \frac{\Im \Pi^L_J(t)}{(t-q^2)^2}
  \\
  \chi^T_J(q^2) &= \frac{\partial \Pi^T_J}{\partial q^2}
  = \frac{1}{\pi} \int_0^\infty dt \frac{\Im \Pi^T_J(t)}{(t-q^2)^2}  
\end{align}
Let us introduce the K\"allen-Lehmann spectral decomposition by
inserting a complete set of states $X$ into the two point function.
\begin{align}
 & (q^\mu q^\nu - q^2 g^{\mu\nu}) \Im \Pi^T_J (q^2)
  + g^{\mu\nu} \Im \Pi^L_J (q^2)
  \CL
  & = \frac{1}{2} \sum_{X} (2\pi)^4 \delta^{4}(q-p_X)
  \langle 0 | J^\mu(0) | X \rangle
  \langle X| [J^\nu(0)]^\dagger | 0 \rangle
  \label{eq:spec-decomp}
\end{align}
where the sum includes an integral over the phase space allowed to
each state $X$ which has the same quantum number as the current $J$.
The positivity of $\Im \Pi^T_J (q^2)$ and $\Im \Pi^L_J(q^2)$ follows
from Eq.~\eqref{eq:spec-decomp} \cite{Bourrely:1980gp}.
In other words,
\begin{align}
  \Big[ (q^\mu q^\nu - q^2 g^{\mu\nu}) \Im \Pi^T_J (q^2)
    + g^{\mu\nu} \Im \Pi^L_J (q^2) \Big] \xi_\mu \xi^*_\nu
  \ge 0
\end{align}
for any complex 4-vector $\xi_\mu$.
This implies that
\begin{align}
  \Im \Pi^T_J (q^2) \ge 0
  \\
  \Im \Pi^L_J (q^2) \ge 0
\end{align}

Let us consider $\Im \Pi^{ii}_J (q^2)$ in Eq.~\eqref{eq:spec-decomp}
(no sum in $i$ index).
\begin{align}
  \Im\Pi^{ii}_J (q^2) =& \frac{1}{2} \int
  \frac{d^3p_1 d^3p_2}{(2\pi)^3 4 E_1 E_2}
  \delta^4( q - p_1 - p_2)
  \CL
  & \times \sum_{\text{pol}} \langle 0 | J^i | H_b(p_1) H_c(p_2) \rangle
  \CL
  & \times \langle H_b(p_1) H_c(p_2) | [J^i]^\dagger | 0 \rangle
  + \cdots \,,
  \label{eq:dr-1}
\end{align}
where the sum is over polarizations of $H_b$ and $H_c$ states, and the
ellipsis denotes strictly positive contributions from the higher
resonances and multi-particle states (three-body or higher multi-body
states).
Here, we assume that $H_b = B, B^*$ meson states, and $H_c = D, D^*$ meson
states.
Since the right-hand side (RHS) of Eq.~\eqref{eq:dr-1} is a sum of
positive contributions, we can obtain the following simple inequality.
\begin{align}
  \Im \Pi^{ii}_J (t) \ge k(t) |F(t)|^2
  \label{eq:ineq:1}
\end{align}
where $t=q^2$, $k(t)$ is a calculable kinematic function arising from
two-body phase space, and $F(t)$ is the form factor associated with a
specific decay of our interest.
For example, in the case of $\bar{B} \to D^* \ell \bar{\nu}$ decay
channel,
\begin{align}
  k(t) |F(t)|^2 =& \frac{1}{2} \int
  \frac{d^3p_1 d^3p_2}{(2\pi)^3 4 E_1 E_2}
  \delta^4( q - p_1 - p_2)
  \CL
  & \times \sum_{\text{pol}} \langle 0 | J^i | B(p_1) D^*(p_2)\rangle
  \CL
  & \times \langle B(p_1) D^*(p_2) | [J^i]^\dagger | 0 \rangle
\end{align}

At this stage, we need to use the second building block: crossing
symmetry
\cite{Boyd:1995sq}.
Let us define $t_\pm \equiv (M_{H_b} \pm M_{H_c})^2$.
The crossing symmetry insures that the $\langle 0 | J^i | H_b
H_c\rangle$ amplitude which shows up in pair production of the $H_b$
and $H_c$ mesons from a virtual $W$ boson shares the same form factor
$F(t)$ as the $\langle \bar{H}_c | J^i | H_b \rangle$ amplitude, while
we can connect the pair production region $t_+ \le t < \infty$ with
the semi-leptonic region $m_\ell^2 \le t \le t_-$ through the analytic
continuation (the third building block).

Let us define the hadronic moments $\chi_J^{(n)}$ as in Ref.~\cite{
Boyd:1995sq, Boyd:1997qw},
\begin{align}
  \chi_J &\equiv \frac{1}{2} \left.
  \frac{\partial^2 \Pi^{ii}_J}{\partial^2 q^2}
  \right|_{q^2 = 0}
  \CL
  &= \bigg[ \frac{\partial \Pi^T_J}{\partial q^2}
    - \frac{1}{2}\frac{\partial^2 \Pi^L_J}{\partial^2 q^2}
    \bigg]_{q^2 = 0}
  \CL
  &= \Big[ \chi^T_J(q^2) - \frac{1}{2}\frac{\partial \chi^L_J}{\partial q^2}
    \Big]_{q^2=0}
  \CL
  &= \frac{1}{\pi} \int_0^\infty dt
  \left. \frac{\Im \Pi^{ii}_J(t)}{(t-q^2)^3} \right|_{q^2=0}
  \\
  \chi_J^{(n)} &\equiv \frac{1}{\Gamma(n+3)}
  \left.
  \frac{\partial^{n+2} \Pi^{ii}_J}{\partial^{n+2} q^2}
  \right|_{q^2 = 0}
  \CL
  &= \frac{1}{\pi} \int_0^\infty dt
  \left. \frac{\Im \Pi^{ii}_J(t)}{(t-q^2)^{n+3}} \right|_{q^2=0}
  \label{eq:moment:n}
\end{align}
where $\chi_J^{(0)} = \chi_J$.
Hence, from the inequality in Eq.~\eqref{eq:ineq:1},
we can obtain the following inequality:
\begin{align}
  \chi_J^{(n)} \ge \frac{1}{\pi} \int^{\infty}_{t_+} dt
  \frac{k(t) |F(t)|^2}{t^{n+3}}
  \label{eq:ineq:2}
\end{align}
where $t_+$ represents the pair production threshold.

At this stage, we need to introduce a key idea of quark-hadron duality
in QCD sum rules which claims that the hadronic moments $\chi_J^{(n)}$
can be calculated in perturbative QCD at $q^2=0$ \cite{Boyd:1997qw}.
Then, we can rewrite the inequality in Eq.~\eqref{eq:ineq:2} as
follows,
\begin{align}
  \frac{1}{\pi} \int^{\infty}_{t_+} dt
  | h^{(n)}(t) F(t) |^2 \le 1\,,
  \label{eq:ineq:3}
\end{align}
where
\begin{align}
  [h^{(n)}(t)]^2 &= \frac{k(t)}{t^{n+3} \chi_J^{(n)}} \ge 0\,.  
\end{align}
The inequality in Eq.~\eqref{eq:ineq:3} imposes an upper bound on the
form factor $F(t)$ in the pair production region ($t_+ \le t <
\infty$).

In order to turn Eq.~\eqref{eq:ineq:3} into a constraint in the
semileptonic region ($m_\ell^2 \le t \le t_-$), we need to use the
third building block: analyticity which allows us to extend the
analytic region of the integrand to the region below the
pair-production threshold ($t<t_+$).
To do this, it is convenient to introduce a conformal mapping function
\begin{align}
  z(t,t_s) &= \frac{ \sqrt{t_+ -t} - \sqrt{t_+-t_s} }{\sqrt{t_+ -t} +
    \sqrt{t_+-t_s} }
\end{align}
$z$ is real for $t < t_+$, $z=-1$ at $t=t_+$, zero at $t = t_s$, and a
U(1) phase $z = e^{i\theta}$ for $t > t_+$.
$t_s \ge t > -\infty$ maps into $0 \le z < 1$ along the real axis.
The upper contour of $t_+ \le t < \infty$ along the real axis maps
into the upper half of a unit circle ($ \pi \ge \theta > 0 $ for $z =
e^{i\theta}$).
Similarly, the lower contour of $t_+ \le t < \infty$ along the real
axis maps into the lower half of a unit circle ($\pi \le \theta < 2\pi$
for $z = e^{i\theta}$).

All the poles in the integrand of Eq.~\eqref{eq:ineq:3} can be removed
by multiplying by various powers of $z(t,t_s)$, if we know the positions
$t_s$ of the sub-threshold poles in $F(t)$ and $h^{(n)}(t)$.
Each pole has a distinct value of $t_s$, and the product $z(t,t_{s1})
z(t,t_{s2}) z(t,t_{s3}) \cdots$ can remove all of them.
For example, such poles include the contribution of $B_c$ resonances
to the form factor $F(t)$ as well as singularities in the kinematic
function $h^{(n)}(t)$.

Once we determine the pole positions phenomenologically, we can
rewrite the inequality in Eq.~\eqref{eq:ineq:3} as follows,
\begin{align}
  \frac{1}{\pi} \int^{\infty}_{t_+} dt
  \left| \frac{dz(t,t_0)}{dt}\right|
  | \phi(t,t_0) P(t) F(t) |^2 \le 1\,,
  \label{eq:ineq:4}
\end{align}
where the outer function $\phi$ is
\begin{align}
  \phi(t,t_0) &= \tilde{P}(t)
  \frac{h^{(n)}(t)}{ \sqrt{ \left| \dfrac{dz(t,t_0)}{dt} \right| } }
\end{align}
The factor $\tilde{P}(t)$ is a product of $z(t,t_s)$'s and
$\sqrt{z(t,t_s)}$'s such that $t_s$ is chosen to remove the
sub-threshold poles and branch cuts in the kinematic function
$h^{(n)}(t)$.
The Blaschke factor $P(t)$ is
\begin{align}
  P(t) &\equiv
  \prod_{i=1}^{N} \frac{ z - z_{P_i} }{ 1 - z z^*_{P_i} }
  = \prod_{i=1}^{N} \frac{ z - z_{P_i} } { 1 - z z_{P_i} }
  \\
  z_{P_i} &\equiv z(t_{P_i}, t_-) =
  \frac{ \sqrt{ t_+ - t_{P_i} } - \sqrt{t_+ - t_-} }
    { \sqrt{ t_+ - t_{P_i} } + \sqrt{t_+ - t_-} } 
\end{align}
where $t_{P_i} = M_{P_i}^2$ represents the pole positions of $F(t)$
below the threshold ($t_{P_i} < t_+$), and $N$ is the number of the
sub-threshold poles in $F(t)$.
Here, note that $z_{P_i}$ is real ($z^*_{P_i} = z_{P_i}$) for the
sub-threshold poles.
In addition, note that $P(t)$ is unimodular ($|P(t)| = 1$), if $z$ is
unimodular ($|z| = 1$).
We have a full freedom to choose $t_0$.
Here, we set $t_0 = t_-$ for convenience and simplicity, and without
loss of generality.\footnote{For other choices of $t_0$, refer to
  Ref.~\cite{ Boyd:1997qw}.}

Since $\phi(t,t_0) P(t) F(t)$ is analytic even in the sub-threshold
region, it is possible to expand this in powers of $z(t,t_0)$.
Hence,
\begin{align}
  F(t) &= \frac{1}{ \phi(t,t_0) P(t)}
  \sum_{n=0}^{\infty} a_n z^n(t,t_0) 
\end{align}
This is called the BGL method of form factor parametrization \cite{
Boyd:1997kz}.
In addition, the integral in Eq.~\eqref{eq:ineq:4} can be rewritten as
follows,
\begin{align}
  \frac{1}{\pi} \int^{\infty}_{t_+} dt
  \left| \frac{dz(t,t_0)}{dt}\right| &=
  \frac{1}{2\pi} \int^{2\pi}_0 d\theta
\end{align}
where $z = e^{i\theta}$ above the threshold ($t_+ \le t < \infty$).
Hence, the final version of the inequality after the Fourier analysis
is
\begin{align}
  \sum_{n=0}^{\infty} a_n^2 \le 1 \,.
\end{align}
This is called the unitarity condition (the weak version).\footnote{A
stronger version can be obtained simply by adding more decay channels
in the right-hand side of the inequality in
Eq.~\eqref{eq:ineq:1} \cite{ Bigi:2017njr}.}

For the $\bar{B} \to D^* \ell \bar{\nu}$ decay process that are the
main subject of this paper, $z(t,t_0)$ is in the physical region of
$0 \le z \le 0.056$ for any physical momentum transfer $m_\ell^2 \le
t \le t_-$.
Hence, in practice, it is possible to truncate the expansion after
the first two or three terms.
Since the BGL method does not use any model to constrain $a_n$,
it is model-independent by construction.

%
%

\section{Conversion formula of top pole quark mass to the top $\MSb$
  quark mass}
\label{app:top-mass}


\subsection{\label{ssec:SI}The scale-invariant mass}

We follow the terminology in the literature.  The pole and \msbar\ masses are
related by the ratio of (mass) renormalization factors $z$,
\begin{align}
\frac{m(\mu)}{M} = z(\mu) \equiv \frac{Z_{\mr{OS}}}{Z_{\msbar}}\,,
\end{align}
where $m(\mu)$ is the \msbar\ mass renormalized at scale $\mu$, $M$ is the pole
mass, $Z_{\mr{OS}}$ is the renormalization factor in the on-shell scheme, and
$Z_{\msbar}$ is the renormalization factor in the \msbar\ scheme.  The ratio
$z$ depends on the scale $\mu$ \textit{via} the strong coupling $\alpha_s(\mu)$
and $\ln(\mu/M)$.
The top scale-invariant (SI) mass $\mu_t$ is the \msbar\ mass $m_t(\mu)$ with
the scale $\mu$ set equal to the scale-invariant mass,
\begin{align}
\mu_t \equiv m_t(\mu_t) \,.\label{eq:SI}
\end{align}
Given the perturbative expansion of $z(\mu)$ in powers of $\alpha_s(\mu)$,
Eq.~\eqref{eq:SI} gives the SI mass in terms of $\alpha_s(\mu_t)$ and
$\ln(\mu_t/M_t)$, \textit{i.e.}, in terms of the SI mass and the pole mass.  To
obtain a formula for the SI mass in terms of the pole mass alone, one can
iterate the perturbative expansion.  The result is an expansion of $\mu_t$ in
powers of $\alpha_s(M_t)$.

\subsection{\label{ssec:three}Three-loop result}

For the three-loop conversion, we use Eq.~(16) of Ref.~\cite{hep-ph/0004189}.
For the top mass conversion, this equation is an expansion of the ratio
$z(\mu_t)$ in powers of the coupling $\alpha_s^{(6)}(M_t)$, with six active
quark flavors.  
We obtain the coupling $\alpha_s^{(6)}(M_t)$ by running the five-flavor
coupling $\alpha_s^{(5)}(M_Z)$ at the $Z$-boson mass to the five-flavor
coupling $\alpha_s^{(5)}(M_t)$ at the top pole and then matching across
threshold.  
The parameter $n_l$ is the number of light quarks; for the top mass conversion,
$n_l = 5$.  
The quantity $\Delta(M_i/M)$ is the two-loop mass correction for a
light quark with pole mass $M_i$; $\Delta(0) = 0$.
The three-loop coefficient $z_m^{\mr{SI},(3)}(M)$ can be found by iterating
Eq.~(13) of Ref.~\cite{hep-ph/0004189} and using Eq.~(15).  We have verified
that doing so yields results consistent with
Refs.~\cite{hep-ph/0004189,hep-ph/9912391} and the RunDec3
code~\cite{1703.03751}.  Our result for the top SI mass is 
\begin{align}
\mu_t = 163.82 \pm 1.05\ \mr{GeV}\,,
\end{align}
where the uncertainty is propagated from the uncertainty in the top pole mass,
and all other uncertainties are neglected.  Below we detail the inputs and the
steps of the calculation.

For the running we use Eq.~(5) of Ref.~\cite{hep-ph/0004189}, which is the
four-loop solution to the RGE, expressed in terms of $\ln(\mu/\Lambda)$, where
$\Lambda$ is the QCD scale.  To obtain $\alpha_s^{(5)}(M_t) = 0.107660$, we set
$\mu = M_t = 173.5 \pm 1.1\ \mr{GeV}$~\cite{Patrignani:2016xqp} and fix
$\ln(M_t/\Lambda)$ using Eq.~(4) of Ref.~\cite{hep-ph/0004189}, with
$\alpha_s^{(5)}(M_Z) = 0.1181$~\cite{Patrignani:2016xqp} and $M_Z = 91.1876\
\mr{GeV}$~\cite{Patrignani:2016xqp}.  
We find $\Lambda = 209.78\ \mr{MeV}$, in agreement with Table 2 of
Ref.~\cite{1703.03751}.
To obtain $\alpha_s^{(6)}(M_t) = 0.107714$, we use the decoupling relation in
Eqs.~(19) and (25) of Ref.~\cite{hep-ph/0004189}, which is the three-loop
expansion of $\alpha_s^{(6)}(M_t)$ in powers of $\alpha_s^{(5)}(M_t)$.  

We set all light-quark mass corrections to zero except that for the bottom
quark.  For the bottom quark pole mass, we use Eq.~(71.21) of
Ref.~\cite{Patrignani:2016xqp}, the three-loop expansion of the pole mass, with
$\mu_b = m_b(\mu_b) = 4.18\ \mr{GeV}$ to obtain $M_b = 4.9324\ \mr{GeV}$.  Then
Eq.~(19) of Ref.~\cite{Gray:1990yh} yields $\Delta(M_b/M_t) = 0.0344909$.

We obtain the three-loop coefficient $z_m^{\mr{SI},(3)}(M)$ in terms of the
coefficients in the expansion of $z(\mu)$ in powers of
$\alpha_s(\mu)$,
\begin{align}
  z(\mu) &= 1 + \frac{\alpha_s(\mu)}{\pi} z_1(\mu)
  + \left(\frac{\alpha_s(\mu)}{\pi}\right)^2 z_2(\mu)
  \CL & \qquad
  + \left(\frac{\alpha_s(\mu)}{\pi}\right)^3 z_3(\mu)
\,.\label{eq:zmu3}
\end{align}
The coefficients $z_1(\mu)$, $z_2(\mu)$, and $z_3(\mu)$ are given explicitly in
Eqs.~(13) and (15) of Ref.~\cite{hep-ph/0004189}.  Taylor expanding the
corresponding terms in Eq.~\eqref{eq:zmu3} about the pole mass $M$, using
the definition of the beta function, setting $m(\mu) = \mu$, and iteratively
solving the result for $\mu/M$ yields
\begin{align}
  \frac{\mu}{M} &= 1 + \frac{\alpha_s(M)}{\pi} z_m^{\mr{SI},(1)}(M)
  + \left(\frac{\alpha_s(M)}{\pi}\right)^2 z_m^{\mr{SI},(2)}(M)
  \CL & \qquad
  + \left(\frac{\alpha_s(M)}{\pi}\right)^3 z_m^{\mr{SI},(3)}(M)\,,
\end{align}
where
\begin{align}
z_m^{\mr{SI},(1)}(M) &= z_1(M)\,,\\
z_m^{\mr{SI},(2)}(M) &= z_2(M) + z_1(M)\, M z_1^{\prime}(M) \,,\\
z_m^{\mr{SI},(3)}(M) &= z_3(M) + z_2(M)\,Mz_1^{\prime}(M) +
z_1(M)\,Mz_2^{\prime}(M)
\CL
&- 2\beta_0 z_1^2(M) + z_1(M)(Mz_1^{\prime}(M))^2
\CL
&+ \ohf (z_1(M))^2\,M^2 z_1^{\prime\prime}(M)\,.
\end{align}
Primes denote derivatives with respect to the scale $\mu$, all coefficients are
evaluated at $\mu = M$, the leading order beta-function coefficient is that for
the six-flavor coupling, and to obtain agreement with the literature, we
neglect the two-loop fermion mass correction when calculating the three-loop
coefficient.  
%

\subsection{\label{ssec:four}Four-loop result}

For the four-loop conversion, we consider the generalization of Eq.~(16) of
Ref.~\cite{hep-ph/0004189} to four loops.  We use five-loop
running~\cite{1606.08659} and four-loop matching~\cite{hep-ph/0512060} to
obtain the coupling $\alpha_s^{(6)}(M_t)$.  For the four-loop coefficient
$z_m^{\mr{SI},(4)}(M)$, we verify that the numerical expression in the RunDec3
code~\cite{1703.03751} agrees with the literature~\cite{1606.06754}.  The
three-loop fermion mass correction is known to be somewhat larger than the
two-loop correction~\cite{1703.03751,0708.1729}, but we neglect it.  Our result
for the SI mass is
\begin{align}
\mu_t = 163.65 \pm 1.05\ \mr{GeV}\,,
\end{align}
where the uncertainty is again that propagated from the pole mass.  Again, we
neglect all other sources of error.
Below we provide details.

The inputs are the same as for the three-loop calculation.  The extension of
Eqs.~(4) and (5) of Ref.~\cite{hep-ph/0004189} to five-loop order are in the
RunDec3 code~\cite{1703.03751}.
%
%
For the QCD scale, we find $\Lambda = 209.80\ \mr{MeV}$, in agreement with
Table 2 of Ref.~\cite{1703.03751}, and for the five-flavor coupling, we find
$\alpha_s^{(5)}(M_t) = 0.107643$.  We match the coupling across threshold using
Eq.~(19) of Ref.~\cite{hep-ph/0004189} with the decoupling factor from
Eqs.~(54,59-63,19,20) of Ref.~\cite{hep-ph/0512060}.  We find
$\alpha_s^{(6)}(M_t) = 0.107703$.

To compare the coefficient $z_m^{\mr{SI},(4)}(M)$ from the RunDec3 code with
the literature, we begin with $z_4(M)$, which enters the expansion of $z(M)$ at
four loops (cf. Eq.~\eqref{eq:zmu3}),
\begin{align}
z(M) &= 1 + \frac{\alpha_s(M)}{\pi} z_1(M) + \left(\frac{\alpha_s(M)}{\pi}\right)^2 z_2(M) \\
&+ \left(\frac{\alpha_s(M)}{\pi}\right)^3 z_3(M) + \left(\frac{\alpha_s(M)}{\pi}\right)^4 z_4(M)
\,.\label{eq:zmu4} \nonumber
\end{align}
We obtain a numerical result for $z_4(M)$ from Eqs.~(15) and (23)
of Ref.~\cite{1606.06754}.  This result agrees with that in the RunDec3 code.
We then relate $z_4(M)$ to $z_m^{\mr{SI},(4)}(M)$ by iterating the expansion of
$z(\mu)$ for $m(\mu) = \mu$ to obtain the expansion of $\mu/M$ in powers of
$\alpha_s(M)$.  We have done this calculation twice, once numerically and once
analytically.  The relation between $z_4(M)$ and $z_m^{\mr{SI},(4)}(M)$ is
\begin{widetext}
\begin{align}
z_m^{\mr{SI},(4)}(M) &= z_4(M) + z_1(M)\,M^2z_1^{\prime\prime}(M) \left( z_2(M) + z_1(M)\,Mz_1^{\prime}(M) \right) \\
&+ (z_1(M))^2\left( \ohf M^2z_2^{\prime\prime}(M) - 2\beta_0\,Mz_1^{\prime}(M) + \beta_0 z_1(M)\right) \nonumber \\
&+ \tfrac{1}{6}(z_1(M))^3\,M^3z_1^{\prime\prime\prime}(M) + z_1(M)\left( Mz_3^{\prime}(M) - 4\beta_0 z_2(M)\right)\nonumber \\
&+ \left( z_2(M) + z_1(M)\,Mz_1^{\prime}(M) \right)\left( Mz_2^{\prime}(M) - 2 \beta_0 z_1(M) \right)\nonumber \\
&+ Mz_1^{\prime}(M)\big( z_3(M) + \ohf (z_1(M))^2\,M^2z_1^{\prime\prime}(M) 
+ z_1(M)\nonumber \\
&\times \left(Mz_2^{\prime}(M) - 2 \beta_0 z_1(M) \right) 
+ Mz_1^{\prime}(M) \left( z_2(M) + z_1(M)\,Mz_1^{\prime}(M) \right) \big)\nonumber \\
&- 2 \beta_1 (z_1(M))^2 \,.\label{eq:zmu4}\nonumber
\end{align}
\end{widetext}
Both our calculations yield agreement with the result for
$z_m^{\mr{SI},(4)}(M)$ in the RunDec3 code, provided that the coefficient
$z_3(\mu)$ given in Eq.~(13) of Ref.~\cite{hep-ph/0004189} is incorrect:  The
first $(\ln \mu^2/M^2)^2$ appearing in the $n_l^2$ term there should be
$\ln(\mu^2/M^2)$.  This evident typo in Ref.~\cite{hep-ph/0004189} does not
affect the three-loop conversion because no derivatives of $z_3(\mu)$ enter.

\bibliography{ref}
\end{document}